\documentclass[twocolumn, twocolappendix]{aastex631} % , twocolappendix gives 2 red errors if there are other red errors; or 1 orange error if there are only other orange errors

\newcommand{\Mjup}{{\rm M}_{\rm Jup}}
\newcommand{\Mearth}{{\rm M}_{\oplus}}
\newcommand{\Msat}{{\rm M}_{\rm Sat}}
\newcommand{\Mnep}{{\rm M}_{\rm Nep}}

\newcommand{\Mth}{{\rm M}_{\rm th}}

\newcommand{\Msun}{{\rm M}_\odot}

\newcommand{\Lsun}{{\rm L}_\odot}

\newcommand{\mm}{\rm mm}

\newcommand{\pc}{\rm pc}
\newcommand{\mas}{\rm mas}
\newcommand{\K}{\rm K}
\newcommand{\au}{\rm au}
\newcommand{\GHz}{\rm GHz}
\newcommand{\kms}{\rm km \, s^{-1}}
\newcommand{\gcmsqrd}{\rm g \, cm^{-2}}
\newcommand{\gcmcbd}{\rm g \, cm^{-3}}
\newcommand{\cmsqrdg}{\rm cm^{2} \, g^{-1}}

\newcommand{\muJybeam}{\mu {\rm Jy \, bm^{-1}}}
\newcommand{\Jybeam}{\rm Jy \, bm^{-1}}
\newcommand{\Jyarcsec}{\rm Jy \, arcsec^{-2}}

\newcommand{\Mp}{M_{\rm p}}

\newcommand{\rp}{r_{\rm p}}
\newcommand{\aplanet}{a_{\rm p}} % to avoid overloading with \ap, \apl commands..

\newcommand{\Omk}{\Omega_{\rm Kep}}
\newcommand{\Omp}{\Omega_{\rm p}}
\newcommand{\vk}{v_{\rm Kep}}
\newcommand{\Sigg}{\Sigma_{\rm gas}}
\newcommand{\Sigz}{\Sigma_{0}}
\newcommand{\Sigemp}{\Sigma_{\rm no \, planet}}
\newcommand{\Tg}{T_{\rm gas}}
\newcommand{\Sigd}{\Sigma_{\rm dust}}
\newcommand{\Td}{T_{\rm dust}}
\newcommand{\Tdiso}{T_{\rm dust, iso}}
\newcommand{\Tdadi}{T_{\rm dust, adi}}

\newcommand{\Tgadi}{T_{\rm gas, adi}}

\newcommand{\tcool}{t_{\rm cool}}
\newcommand{\rhod}{\rho_{\rm dust}}
\newcommand{\rhog}{\rho_{\rm gas}}
\newcommand{\vth}{v_{\rm th}}

\newcommand{\hp}{h_{\rm p}}
\newcommand{\St}{\rm St}
\newcommand{\Stcrit}{\rm St_{\rm crit}}
\newcommand{\deltaphispiral}{\Delta \phi_{\rm gas \, spiral}}

\newcommand{\agrain}{a_{\rm grain}}

\newcommand{\vrdust}{v_{r,{\rm dust}}}
\newcommand{\tstop}{t_{\rm stop}}
\newcommand{\tdyn}{t_{\rm dyn}}
\newcommand{\tcross}{t_{\rm cross}}
\newcommand{\ttherm}{t_{\rm therm}}
\newcommand{\vdust}{\vec{v}_{\rm dust}}
\newcommand{\vgas}{\vec{v}_{\rm gas}}
\newcommand{\adust}{\vec{a}_{\rm dust}}
\newcommand{\gravity}{\vec{g}}

\newcommand{\Mstar}{M_\star}
\newcommand{\Lstar}{L_\star}

\newcommand{\Inu}{I_\nu}
\newcommand{\taunu}{\tau_\nu}
\newcommand{\tauz}{\tau_{0}}
\newcommand{\Bnu}{B_\nu}

\newcommand{\kappanu}{\kappa_{\nu}}

\newcommand{\nuobs}{\nu_{\rm obs}}
\newcommand{\lambdaobs}{\lambda_{\rm obs}}

\newcommand{\mprot}{m_{\rm p}}
\newcommand{\kB}{k_{\rm B}}

\newcommand{\thetaAR}{\theta_{\rm AR}}
\newcommand{\thetaMRS}{\theta_{\rm MRS}}
\newcommand{\thetaLAS}{\theta_{\rm LAS}}

\newcommand{\D}{\rm d}

\shorttitle{Planet-driven dust spirals with ALMA}
\shortauthors{Speedie et al. 2022}
\graphicspath{{./}{figs_compressed/}}
\begin{document}

\title{Observing planet-driven dust spirals with ALMA}
%\title{Observational Signatures of Planets in Protoplanetary Disks: Observing planet-driven dust spirals with ALMA}
% \newcommand{\rdtext}[1]{{\bf \color{green}[#1]}}
% \newcommand{\rdnote}[1]{{\bf \color{red}[R.D.: #1]}}

\correspondingauthor{JS, RD}
\email{jspeedie@uvic.ca, rbdong@uvic.ca}

\author[0000-0003-3430-3889]{Jessica Speedie}
\affiliation{Department of Physics \& Astronomy, University of Victoria, Victoria, BC, V8P 1A1, Canada}

\author[0000-0002-0364-937X]{Richard A. Booth}
\affiliation{Astrophysics Group, Imperial College London, Prince Consort Road, London SW7 2AZ, UK}

\author[0000-0001-9290-7846]{Ruobing Dong}
\affiliation{Department of Physics \& Astronomy, University of Victoria, Victoria, BC, V8P 1A1, Canada}

\begin{abstract} % limit: 250 words; current: 245 words

ALMA continuum observations of thermal emission from the dust component of protoplanetary disks have revealed an abundance of substructures that may be interpreted as evidence for embedded planets, but planet-driven spiral arms --perhaps one of the most compelling lines of evidence-- have proven comparatively elusive. In this work, we test the capabilities of ALMA to detect the planet-driven spiral signal in continuum emission.
Carrying out hydrodynamic simulations and radiative transfer calculations, we present synthetic Band 7 continuum images for a wide range of disk and observing conditions.
We show that thermal mass planets at tens of au typically drive spirals detectable within a few hours of integration time, and the detectable planet mass may be as low as $\sim$Neptune mass ($0.3 \, \Mth$).
The grains probed by ALMA form spirals morphologically identical to the underlying gas spiral.
The temperature of the dust spiral is crucial in determining its contrast, and spirals are easier to detect in disks with an adiabatic equation of state and longer cooling times.
Resolving the spiral is not necessary for its detection; with the help of residual maps, 
the optimal beam size is a few times the spiral width at a constant noise level.
Finally, we show how the presence of gaps and rings can impair our ability to recognize co-located spirals. Our work demonstrates the planet-finding potential of the current design specification of ALMA, and suggests that observing capability is not the bottleneck in searching for spirals induced by thermal mass planets.

\end{abstract}


%%%%%%%%%%%%%%%%%%%%%%%%%%%%%%%%%%%%%%%%%%%%%%%%%%%%%%%%%%%%%%%%%%%%%%%%%%%%%%%%
%%%%%%%%%%%%%%%%%%%%%%%%%%%%%%%%%%%%%%%%%%%%%%%%%%%%%%%%%%%%%%%%%%%%%%%%%%%%%%%%
\section{Introduction} \label{sec:intro}
%%%%%%%%%%%%%%%%%%%%%%%%%%%%%%%%%%%%%%%%%%%%%%%%%%%%%%%%%%%%%%%%%%%%%%%%%%%%%%%%
%%%%%%%%%%%%%%%%%%%%%%%%%%%%%%%%%%%%%%%%%%%%%%%%%%%%%%%%%%%%%%%%%%%%%%%%%%%%%%%%

Like the wake created by a boat as it moves through water, a planet drives a wake as it orbits in a disk \citep{2002-ogilvie}. The wake then gets wound into a spiral 
by the disk's own Keplerian differential rotation \citep{2018-arzamasskiy}. Planet-driven spiral arms are a well understood natural consequence of the gravitational interaction between the planet and the disk; the analytical theory was established in the 1970's \citep{1978-goldreich, 1979-goldreich, 1980-goldreich} 
and has been repeatedly confirmed by hydrodynamical simulations \citep{2011b-dong, 2011a-dong, 2015-zhu-dong-shocks, 2018a-bae-zhu, 2018b-bae-zhu, 2019-miranda-rafikov}. We therefore expect that within every planet-hosting protoplanetary disk, spiral wakes should also exist.

This theoretical expectation has not translated into an abundance of clear observational detections of planet-driven spirals, however. 
That statement particularly applies to continuum observations of (sub-)mm dust. High angular resolution ALMA continuum observations have shown that gaps and rings are common \cite[e.g.,][]{2018a-huang-dsharp2}, implying an abundance of planets forming in disks \citep{2018-zhang-dsharp7}, 
but to date we have only a handful of detections of continuum spirals. 
Elias 27, IM Lup and WaOph 6 \citep{2018-huang-dsharp3-spirals} exhibit large-scale $m=2$ continuum spirals, but they are not co-located with a gap/cavity, and have not been decisively attributed to an embedded companion \citep{2012-mawet-imlup, 2017-meru-elias227, 2021-paneque-carreno-elias227, 2021-brown-sevilla-waoph6}. A smaller scale, 
single continuum spiral has been observed in MWC 758 \citep{2018-dong-MWC758}; the motion of that disk's related $m=2$ scattered light spirals have been reported as inconsistent with gravitational instability \citep{2020-ren-mwc758} but no associated point-source has yet been detected \citep{2021-boccaletti-MWC758}. Tentative crescents or filaments in association with gaps/rings have been observed in continuum images of V1247 Ori \citep{2017-kraus-v1247} and HD 135344B \citep{2021-casassus-filament} but haven't been confidently classified as spirals. Continuum spirals in disk systems with multiple stars, HD100453 \citep{2020-rosotti-HD100453}, AS 205 N and HT Lup A \citep{2018-kurtovic-dsharp4} have been identified as induced by gravitational interaction with companions, but the companions are not of planetary mass.

Searches for planet-driven dust spirals in continuum observations, and recognizing the signatures that a spiral is planet-driven if it is found, would benefit from a clearer understanding of the following complexities:

(1) It is well understood 
how planets drive spirals in the gas, but it is not necessarily obvious how spirals manifest in the dust. The morphology (i.e., the amplitude, width, and azimuthal location) of dust spirals is determined by how quickly each particle responds to the drag forces exerted by the passing spiral wake \citep{2020-sturm}. The response time of the dust depends on how well it is coupled to the gas, which in turn depends on the grain properties and local gas density \cite[the latter making it a function of radial location and height above the midplane, e.g. Eqn. 1 of][]{2019-veronesi-diskmass}. 
For each observing wavelength and instrument, we need to understand what architecture of dust spiral we are looking for.

(2) Compounding this is the fact that planet-driven spirals are not dust traps. Since the gas spiral wake co-moves with the planet as it orbits the star, the dust experiences the spiral perturbation as a transient phenomenon. As a result, a dust spiral's density amplitude 
cannot exceed 
that of the gas. This is different to the case of dust rings, where dust can accumulate in a stationary (or at least, long-lived) pressure maximum over time \citep{1972-whipple}. In this way, a lower mass planet can still produce readily detectable dust gaps/rings \cite[e.g.,][]{2016-rosotti, 2017-bae, 2017-dong-multiple1, 2018-dong-multiple2}, but a lower mass planet drives proportionally lower amplitude spiral arms \citep{2011a-dong, 2018a-bae-zhu, 2019-miranda-rafikov}. We need to understand what level of contrast a planet-driven spiral can achieve in the dust density for planets in the ``still-forming'' mass range.

(3) How we can actually observe the dust density distribution is, in the case of continuum observations, through its thermal emission -- which then introduces the question of dust temperature and the rate at which the disk cools. The importance of using a realistic treatment of disk thermodynamics in simulations whose purpose is to interpret or predict planet-induced disk substructure is gaining recognition \citep{2019-miranda-rafikov-planet-interpretation}. One's choice of the equation of state has been shown to have significant consequences on the density wave dynamics \citep{2020b-miranda-rafikov}. The rate of cooling, specifically, affects the angular momentum flux across the disk, which modifies the gas spiral density amplitude \citep{2020a-miranda-rafikov, 2020-zhang-zhu-cooling}. The hydrodynamic $P {\rm d}V$ work done on the gas as the spiral pressure wave passes generates a rise in the temperature distribution, forming temperature spirals whose amplitudes can be observationally significant \citep{2021-muley}. For a given disk's thermodynamic properties (and optical depth), we need to understand how the density and temperature spirals combine into the observed quantity of intensity -- for the dust.

(4) Finally, there is the practical consideration of the angular resolution and sensitivity at which we observe. Once we understand the dust spiral morphology and possible intensity contrast, we can establish which observing specifications, and which disks, give us the best chance of detecting planet-driven dust spirals.

In this work, we carry out this experiment for the case of sub-mm dust continuum observations with Band 7 of ALMA. 
Our purpose is to aid searches for planet-driven dust spirals in existing ALMA observations and to inform future observing proposals.

\S\ref{sec:methods} describes our methodology. In \S\ref{sec:results1} we present our results on the important physics: dust-gas coupling (Complexities 1 \& 2) and thermodynamics (Complexity 3). In \S\ref{sec:results2} (Complexity 4) we present synthetic ALMA continuum observations of planet-driven dust spirals for a variety of disk and observing conditions. We discuss our results in \S\ref{sec:discussion} and summarize our findings in \S\ref{sec:conclusions}.

%%%%%%%%%%%%%%%%%%%%%%%%%%%%%%%%%%%%%%%%%%%%%%%%%%%%%%%%%%%%%%%%%%%%%%%%%%%%%%%%
%%%%%%%%%%%%%%%%%%%%%%%%%%%%%%%%%%%%%%%%%%%%%%%%%%%%%%%%%%%%%%%%%%%%%%%%%%%%%%%%
\section{Methods} \label{sec:methods}
%%%%%%%%%%%%%%%%%%%%%%%%%%%%%%%%%%%%%%%%%%%%%%%%%%%%%%%%%%%%%%%%%%%%%%%%%%%%%%%%
%%%%%%%%%%%%%%%%%%%%%%%%%%%%%%%%%%%%%%%%%%%%%%%%%%%%%%%%%%%%%%%%%%%%%%%%%%%%%%%%

\begin{figure*}
\plotone{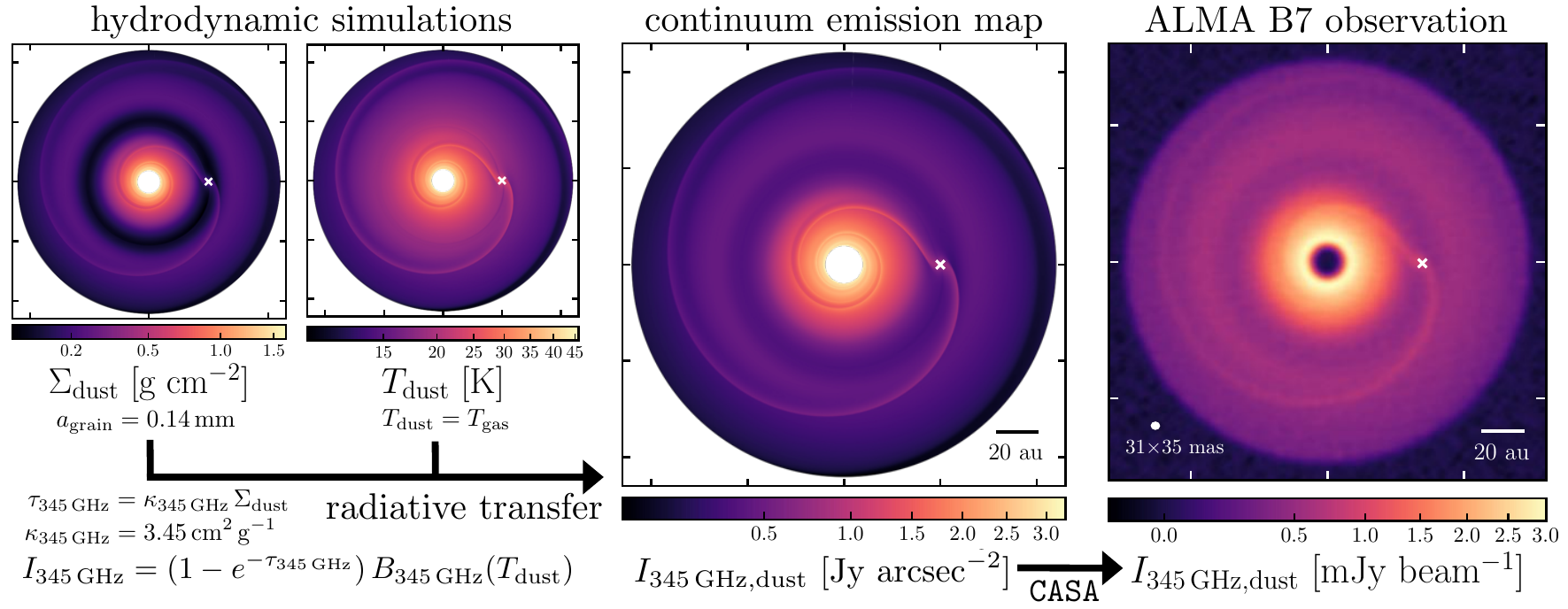}
\caption{The procedure by which we generate synthetic ALMA continuum observations of planet-driven spiral arms. Dust surface density and temperature maps from our hydrodynamic simulations are combined via radiative transfer calculations to create model images of dust thermal emission (emergent intensity), with which we generate continuum observations using \texttt{CASA}. Shown here is the case of a $1 \, \Mth$ planet embedded in an adiabatic ($\beta=10$), marginally optically thick disk ($\tauz=1.0$), observed with the C43-5 + C43-8 configuration pair and a combined 8.0 hrs of on-source time (measured rms noise $9.17 \, \muJybeam$). 
\label{fig:sec2-overview}}
\end{figure*}

We run 2D gas + dust hydrodynamic simulations of disks with different cooling rates, optical depths and embedded planet masses to obtain the dust density and temperature distributions at the disk midplane (\S\ref{subsec:hydrosims}). The dust grain size is fixed to $\agrain = 0.14 \, \mm$ to correspond to Band 7 and the gas surface density is varied to achieve different dust optical depths assuming a fixed dust-to-gas mass ratio of $0.01$. Next, we create synthetic continuum emission images of the resulting dust spirals via radiative transfer calculations (\S\ref{subsec:radtranscalcs}). With those we generate synthetic ALMA observations for a range of integration times and antenna configurations (\S\ref{subsec:almaobs}). Figure \ref{fig:sec2-overview} provides examples of the outputs after each step of our procedure.

%%%%%%%%%%%%%%%%%%%%%%%%%%%%%%%%%%%%%%%%%%%%%%%%%%%%%%%%%%%%%%%%%%%%%%%%%%%%%%%%
\subsection{Hydrodynamic Simulations} \label{subsec:hydrosims}
%%%%%%%%%%%%%%%%%%%%%%%%%%%%%%%%%%%%%%%%%%%%%%%%%%%%%%%%%%%%%%%%%%%%%%%%%%%%%%%%

We perform 2D multi-fluid hydrodynamic simulations with a custom version of the \texttt{FARGO3D} code \citep{2016-benitez-llambay-masset}, modified to compute dust dynamics with the Eulerian implementation described in \citet{2016-rosotti}, which uses the semi-implicit integrator introduced by \citet{2015-booth}. %

We run the simulations in a 2D cylindrical geometry $(r, \phi)$. The units are dimensionless, such that the orbital radius of the planet ($\rp$) is unity, the unit of time is the Keplerian angular velocity $\Omk$ at $r=\rp$, and the unit of mass is that of the central star. The domain extends from $0.1 \, \rp$ to $3.0 \, \rp$ in radius and from $-\pi$ to $\pi$ in azimuth. The grid has a resolution of $N_{r} \times N_{\phi} = 1100 \times 2048$ cells, spaced logarithmically and linearly in the radial and azimuthal directions respectively, for approximately square cells. With this resolution, one scale height at $\rp$ is resolved with 23 cells in both directions, and we resolve the spiral shock fronts.

The code solves the mass, momentum and energy equations of hydrodynamics:
\begin{equation}
    {\partial \Sigg \over \partial t} + \nabla \cdot (\Sigg \vgas) = 0 \, ,
    \label{eqn:hydro-mass}
\end{equation}
\begin{equation}
    {\partial \vgas \over \partial t} + \vgas \cdot \nabla \vgas + {\nabla P \over \Sigg} = \gravity \, ,
    \label{eqn:hydro-momentum}
\end{equation}
\begin{equation}
    {\partial E \over \partial t} + \vgas \cdot \nabla E + {P \over \Sigg} \nabla \cdot \vgas = 0 \, ,
    \label{eqn:hydro-energy}
\end{equation}

where $\Sigg$, $\vgas$, $P$ and $E$ are the gas surface density, velocity, pressure, and internal energy per unit area, and $\gravity$ is the gravity term. 

We solve the energy equation for the gas and simultaneously evolve the dust\footnote{To our knowledge, ours is the first work using this custom version of \texttt{FARGO3D} to do so.}. The dust is treated as a pressureless fluid and evolves according to linear drag forces from the gas, in addition to gravity and diffusion. The dust velocity is given by
\begin{equation}
    {d \vdust \over d t} + \vdust \cdot \nabla \vdust = - {1 \over \tstop} \Big( \vdust - \vgas(t) \Big) + \adust \, ,
    \label{eqn:dust-velocity}
\end{equation}
where $\tstop$ is the stopping time of the dust and $\adust$ is the non-drag acceleration. 
We focus on dust with dimensionless stopping time $\tstop \, \Omk \leq 1$, for which the fluid approximation is reasonable \citep{2004-garaud}.
Throughout our simulations the dust-to-gas ratio never approaches unity, thus the back reaction from the dust onto the gas is unimportant and ignored. We include dust diffusion, and the Schmidt number (the ratio of the $\alpha$-viscosity $\nu$ to the dust diffusion coefficient $D$) is set to ${\rm Sc} = \nu / D = 1$.

In order to explore the observability of planet-driven dust spirals in a diversity of disk conditions, we vary the gas equation of state to be locally isothermal or adiabatic. The relationship between the gas pressure, density and temperature naturally affects the gas spiral, which in turn we expect to affect the resultant dust spiral via dust-gas coupling. 

The adiabatic equation of state is $P = (\gamma - 1) \, E$, 
and the gas temperature is
\begin{equation}
    \Tgadi = {\mu \, \mprot \over \kB} \, {(\gamma -1) \, E \over \Sigg} \, .
    \label{eqn:adi-eos}
\end{equation}
Here, $\mu=2.3$ is the chemical potential for fully molecular gas of cosmic composition (mix of H and He), $\mprot$ is the mass of a proton, $\kB$ is the Boltzmann constant and $\gamma$ is the adiabatic index. In this work, we assume $\gamma = 5/3$, appropriate for a composition of H$_{2}$ at a low temperature (tens of Kelvin). 
We perform the adiabatic simulations with a simple cooling prescription, such that the gas temperature is relaxed towards its initial state on a timescale controlled by the parameter $\beta$:
\begin{equation}
    {d E(t) \over d t} = - {\Omega \over \beta} \Big( E(t) - E_{0} \Big) \,
    \label{eqn:energy-cooling-diff} 
\end{equation}
where $\Omega$ is the local angular velocity. Previous work has shown that simulations with an adiabatic equation of state and short cooling times ($\beta \lesssim 10^{-1}$) yield small planet-induced temperature perturbations and are very similar to simulations with an isothermal equation of state \citep{2020a-miranda-rafikov, 2020b-miranda-rafikov, 2020-zhang-zhu-cooling, 2021-muley}. 
Analytic expressions derived 
assuming conventional dust properties indicate that the cooling timescale can vary dramatically within a single disk at different radii, and typical values are $\tcool= \beta \, \Omega^{-1}(r) \sim 20$ at $r=10 \, \au$ and $\tcool \sim 0.02$ at $r=100 \, \au$ \citep[Eqn. 39 of][]{2020-zhang-zhu-cooling}. Thus in our experiments we explore $\beta$ values of 0 (isothermal) and 10.
We note that while the code does not add the energy dissipated via physical viscosity (e.g. $\alpha$ viscosity) to the internal energy of the gas, the heating due to the artificial viscosity (for handling shocks) has been included\footnote{Viscous heating is also missing in the public version of \texttt{FARGO3D} \citep{2016-benitez-llambay-masset}.}. Viscous heating is likely unimportant for our purposes as we are modelling Class II disks with planets at tens of au.

We initialize the aspect ratio in the disk as
\begin{equation}
    h(r) = H/r = \hp \, r^{f} \, ,
    \label{eqn:h-fargo3d}
\end{equation}
where we choose a flaring index of $f=1/4$ and value of $h$ at the location of the planet $\hp = 0.07$ (see \S\ref{subsec:radtranscalcs} for physical motivation). We use the $\alpha$ viscosity prescription of \citep{1973-shakura} and assume the conventional $\alpha = 10^{-3}$. Varying the viscosity does not impact the spiral arms \citep{2017-dong-fung}, while it makes an impact on the gap depth \citep{2014-fung-gaps}.

We tailor the setup of the dust component of our hydrodynamic simulations toward the end goal of ALMA Band 7 continuum observations by fixing the dust grain size. For the observing wavelength $\lambdaobs = 0.87 \, \mm $ (observing frequency $\nuobs = 345\, \GHz$), we assume we probe thermal emission from dust particles of size \citep{2015-kataoka, 2019-pavlyuchenkov}
\begin{equation}
    \agrain \approx {\lambdaobs \over 2 \pi} \, ,
    \label{eqn:agrain-lambdaobs}
\end{equation}
giving a dust grain size of $\agrain = 0.14 \, \mm$. This means that the dust Stokes number, 
\begin{equation}
    \St = \tstop \, \Omk = {\pi \over 2} \, {\agrain \, \, \rhod \over \Sigg} \, ,
    \label{eqn:stokes-number}
\end{equation}
varies in space and time inversely to the gas surface density. We assume $\rhod = 1.2 \, \gcmcbd$ for the bulk grain density \cite[in broad agreement with][]{2018-birnstiel-dsharp5}. The initial gas surface density distribution is assumed to follow a power law
\begin{equation}
    \Sigg(r) = \Sigz \, \Big( {r \over \rp} \Big)^{-1} \, ,
    \label{eqn:sigma0-fargo3d}
\end{equation}
where $\Sigz$ is the initial gas surface density at $r=\rp$.
%and we set $p=1$. 
Since the simulations are scale-free and we ignore dust feedback, the normalization factor $\Sigz$ is arbitrary and we can use $\Sigz$ to scale our simulation results during post-processing to the physical $\Sigg$ and $\St$ that matches the $\agrain$ we desire. More on this in \S\ref{subsec:radtranscalcs}.

We run our simulations with three different planet masses: $\Mp = 0.3$, $1.0$ and $3.0 \, \, \Mth$ (where the thermal mass is $\Mth = h^3 \Mstar$), corresponding to planet-star mass ratios of $q = 1.03 \times 10^{-4}$, $3.43 \times 10^{-4}$ and $1.03 \times 10^{-3}$. With an aspect ratio of $\hp = 0.07$ and a stellar mass of $\Mstar = 0.8 \, \Msun$, those masses equate to $1.6 \, \Mnep$, $0.96 \, \Msat$ and $0.86 \, \Mjup$ respectively. We keep the planet on a fixed circular orbit and include the indirect term that compensates for the displacement between the simulation grid origin and the star-planet center of mass. 
We simulate to 1500 orbits so that gaps and rings, a commonly observed category of disk substructure, have fully formed \cite[e.g. Fig. 1,][]{2016-fung-gaps}.

In addition to the simulations that we use to generate ALMA observations (presented in \S\ref{sec:results2}), we run a separate shorter set (15 orbits) for the purposes of deepening our understanding the effect of the cooling time and dust-gas coupling on the spiral's intrinsic properties (presented only in \S\ref{subsec:dustgas-coupling} and \S\ref{subsec:temperature-spirals}). In this set, we vary the cooling time from $\beta=10^{-3}$ to $10^{2}$ in factors of 10. Following \citet{2020-sturm}, we evolve 40 species of dust with a spatially and temporally constant Stokes number logarithmically spaced between $\St = 10^{-4}$ and $1.0$. Otherwise, the simulation setup is the same.

\subsubsection{Boundary Conditions} \label{subsubsec:boundcond}

For the gas component, we use closed boundaries where the surface density and azimuthal velocity fields are scaled using the closest active cell, the radial velocity field is mirrored, and the energy field is extrapolated symmetrically. For the dust component, we use open/inflow boundary conditions. Like \citet{2020-sturm}, we set the radial velocity at the boundaries to the radial drift velocity \cite[an extrapolation based on Eqns. 23-26 of][]{2002-takeuchi-lin}:
\begin{equation}
    \vrdust = {\eta \, \vk \over \St + \St^{-1}} \propto \Big( {H \over r} \Big)^2 \, \vk \propto r^{2f - 1/2}
    \label{eqn:vrdust-sturm-bc}
\end{equation}
where $\vk$ is the Keplerian velocity and $\eta = \big( {H \over r} \big)^2 \, \frac{d \log (P)}{d \log(r)}$. These radial velocity boundary conditions are desirable for us because they negate the effect of radial drift accumulating over time, meaning we avoid any dust pile-up at the inner boundary or depletion of dust in the outer disk that might impinge on the expression of the planet-driven spiral. 

At the radial boundaries, we employ wave damping zones \citep{2006-deval-borro} to minimize wave reflections, where the damping zones' inner and outer edges have a Keplerian orbital frequency ratio of 2/3 \cite[Eqns. A3 \& A4 of][]{2019-mcnally}. For our domain, the inner and outer edges of the damping regions work out to $0.131 \, \rp$ and $2.289 \, \rp$. We apply damping to each of the density, azimuthal and radial velocity fields, with a local damping timescale of $1/[30 \, \Omk(r)]$. 

%%%%%%%%%%%%%%%%%%%%%%%%%%%%%%%%%%%%%%%%%%%%%%%%%%%%%%%%%%%%%%%%%%%%%%%%%%%%%%%%
\subsection{Radiative Transfer Calculations} \label{subsec:radtranscalcs}
%%%%%%%%%%%%%%%%%%%%%%%%%%%%%%%%%%%%%%%%%%%%%%%%%%%%%%%%%%%%%%%%%%%%%%%%%%%%%%%%

Prior to any post-processing, we radially truncate the disk to extend from $0.2 \, \rp$ to $2.2 \, \rp$ in order to remove the damped zones. We model our choice of physical disk parameters after the disks in the DSHARP survey, setting the stellar mass and stellar luminosity to the median of the sample, $\Mstar = 0.8 \, \Msun$ and $\Lstar = 1.5 \, \Lsun$ \cite[Table 1 of][]{2018-andrews-dsharp1}, and the orbital radius of the planet to be roughly coincident with the radial location where many of the DSHARP gaps and rings are found, $\aplanet = 50 \, \au$ \cite[Fig. 7 of][]{2018a-huang-dsharp2}. The disk thus extends from $10$ to $110 \, \au$. We place the disk at a distance of $d = 140 \, \pc$ and assume it has zero inclination (i.e. is face-on), giving the disk diameter ($220\, \au$) an angular size of $1.57 \arcsec$. 

A key point we emphasize in this work is that both the dust surface density and dust temperature contribute to the emergent dust intensity, $\Inu$, the observed quantity in ALMA continuum images. We calculate $\Inu$ as
\begin{equation}
    \Inu = \Bnu(\Td) \cdot (1 - e^{-\taunu}) \, ,
    \label{eqn:radiative-transfer}
\end{equation}
where $\Bnu(T)$ is the Planck function, $\Td$ is the dust temperature, and $\taunu$ is the disk optical depth. Throughout, our observing frequency is $\nu = 345 \, \GHz$. The dust temperature contributes via the first factor, $\Bnu(\Td)$, and the dust surface density via the second, $(1 - e^{-\taunu})$.

In the first factor, $\Bnu(\Td)$, we use a different dust temperature distribution for each equation of state. For the adiabatic simulations, we convert the \texttt{FARGO3D} output (gas) energy field into temperature in units of Kelvin, and assume thermal equilibrium between gas and dust such that $\Tdadi = \Tg$. We justify this assumption in \S\ref{sec:app:Tdust}.

For the isothermal simulations (which don't solve the energy equation), we create an axisymmetric dust temperature map using Eqn. B2 of \citet{2018b-dong}:
\begin{equation}
    \Tdiso = 13.37 \, \Big({r \over 100 {\rm \, au}} \Big)^{-1/2} \,
    \label{eqn:Tdust-iso-dong2018b}
\end{equation}
which is the temperature profile consistent with disks heated by a central star of luminosity $\Lstar=1.5\, \Lsun$ and aspect ratio of Eqn. \ref{eqn:h-fargo3d}. At a radius of $\aplanet = 50$ au, Eqn. \ref{eqn:Tdust-iso-dong2018b} yields a temperature of $18.9 \, \K$, a sound speed of $0.26 \, \kms$, a physical scale height of $3.44 \, \au$ and an aspect ratio just under 0.07 (hence our selection of $\hp = 0.07$ in \S\ref{subsec:hydrosims}). 
A physical disk may have a vertical temperature gradient \cite[e.g.,][]{2020-rosotti-HD100453}, but (sub-)mm-sized grains are expected to settle to the disk midplane and adopt the temperature there.

In the second factor, $(1 - e^{-\taunu})$, the disk optical depth is
\begin{equation}
    \taunu = \kappanu \cdot \Sigd \, ,
    \label{eqn:optical-depth}
\end{equation}
where $\Sigd$ is the dust surface density output from \texttt{FARGO3D} and $\kappa_{345 \, \GHz} = 3.45 \, \cmsqrdg$ \citep{1990-beckwith}. 

We vary the disk optical depth by initializing the hydro simulations with a fixed dust-to-gas ratio of $0.01$ and different $\Sigz$ (Eqn. \ref{eqn:sigma0-fargo3d}). The dust-to-gas ratio then evolves at each radii in the simulations.
Throughout the paper, we use the term $\tauz$ to represent the optical depth at $\nu = 345 \, \GHz$ and $r=\rp$ at the beginning of our simulations, and we construct our simulation input parameters to give $\tauz = 0.1$, $0.3$, $1.0$, and $3.0$. Of course, a particle of fixed grain size embedded in a higher density gas disk will be better coupled, and we account for this by setting up the initial dust Stokes number accordingly. For example, the dust in our $\tauz = 0.1$ disk has an initial Stokes number of $\St(\rp) = 9 \times 10^{-3}$ via $\Sigz = 2.89 \, \gcmsqrd$, and the optically thickest $\tauz=3.0$ disk has initial $\St(\rp) = 3 \times 10^{-4}$ via $\Sigz = 86.9 \, \gcmsqrd$. 
For a reference comparison between the $\tauz = 0.1$, $0.3$, $1.0$ and $3.0$ disks, Figure \ref{fig:app_diskquantities} in \S\ref{sec:app:sec2-supplementary} provides radial profiles of $\Sigg$, $\Sigd$, $\St$, $\Td$, $\tau_{345 \, \GHz}$ and $I_{345 \, \GHz}$ for our adiabatic ($\beta=10$) disk with a $1.0 \, \Mth$ embedded planet.

%%%%%%%%%%%%%%%%%%%%%%%%%%%%%%%%%%%%%%%%%%%%%%%%%%%%%%%%%%%%%%%%%%%%%%%%%%%%%%%%
\subsection{Synthetic ALMA Observations} \label{subsec:almaobs}
%%%%%%%%%%%%%%%%%%%%%%%%%%%%%%%%%%%%%%%%%%%%%%%%%%%%%%%%%%%%%%%%%%%%%%%%%%%%%%%%

To explore ALMA's capability to detect planet-driven dust spirals, we generate synthetic Band 7 continuum observations for a range of integration times and antenna configurations with the \texttt{CASA} 
software package \citep{2007-mcmullin-casa}. Our choice of observing band strikes a balance between angular resolution (favouring shorter wavelengths), signal-to-noise ratio (favouring longer wavelengths), feasibility (disfavouring Bands 8-10), and popularity (favouring Bands 6 or 7).

We observe the disk with both a compact and extended 12m array configuration to simultaneously achieve a high angular resolution and a large maximum recoverable scale. We choose the configuration pairs C43-4 + C43-7, C43-5 + C43-8 and C43-6 + C43-9 following the Cycle 8 Proposer's Guides. 
The maximum recoverable scales of the compact configurations C43-4, C43-5 and C43-6 are $\thetaMRS = 3.3 \arcsec$,  $1.9\arcsec$, and $1.2\arcsec$, respectively, so the C43-6 configuration does not quite cover emission on the angular scale of the disk, $\thetaLAS=1.57 \, \arcsec$. Observing with a compact configuration in addition to an extended one requires $20-22\%$ more on-source time, but we found that doing so improves uv-sampling, reducing long baseline artifacts that exist with the high angular resolution configurations, giving the combined image overall higher quality; see Fig. \ref{fig:app_singleconfig} in \S\ref{sec:app:sec4-supplementary} for an illustration.

We set up the integration time to target a requested continuum sensitivity of $10$, $15$, $20$, $25$, $30$ and $35 \, \muJybeam$. As per the Sensitivity Calculator in the ALMA Cycle 8 OT, these requested sensitivities correspond to a combined on-source time (i.e., summed compact and extended configuration on-source time) of roughly 8.0 hrs, 3.5 hrs, 2.0 hrs, 1.3 hrs, 55 min, and 40 min, depending on the configuration pair. Table \ref{tab:sensitivity-obstime} in \S\ref{sec:app:sec4-supplementary} provides the individual and combined on-source times, as well as the total time including overheads, for each requested sensitivity and antenna configuration pair.

We generate measurement sets for the compact and extended configurations using the \texttt{simobserve} tool. The disk is assumed to have RA and Declination J2000 19h00m00 -40d00m00. We set the continuum bandwidth at $345 \, \GHz$ to the full available $7.5 \, \GHz$, and adopt the default choices of a precipitable water vapour level of $0.913 \, \mm$ and ambient ground temperature of $269$ K.
We concatenate the measurement sets from both configurations for each model and clean them simultaneously with \texttt{tclean}, and then generate the noisy images with \texttt{simanalyze}. \texttt{CLEAN} images are created using a briggs weighting and a robust factor of 0.5. We clean to a threshold of $3 \times$ the requested sensitivity. After cleaning, the measured rms noise in all our images presented in \S\ref{sec:results2} is $\sim$80-95\% of the requested sensitivity (with cleaning affecting those images with poorer uv-coverage more greatly).

%%%%%%%%%%%%%%%%%%%%%%%%%%%%%%%%%%%%%%%%%%%%%%%%%%%%%%%%%%%%%%%%%%%%%%%%%%%%%%%%
%%%%%%%%%%%%%%%%%%%%%%%%%%%%%%%%%%%%%%%%%%%%%%%%%%%%%%%%%%%%%%%%%%%%%%%%%%%%%%%%
\section{Important Physics for the Observability of Planet-driven Dust spirals} \label{sec:results1}
%%%%%%%%%%%%%%%%%%%%%%%%%%%%%%%%%%%%%%%%%%%%%%%%%%%%%%%%%%%%%%%%%%%%%%%%%%%%%%%%
%%%%%%%%%%%%%%%%%%%%%%%%%%%%%%%%%%%%%%%%%%%%%%%%%%%%%%%%%%%%%%%%%%%%%%%%%%%%%%%%

We employ two different metrics to quantify the characteristics of planet-driven spirals: (1) ``perturbation'', and (2) ``contrast''. We calculate the perturbation in $X$ disk quantity relative to an unperturbed disk as
\begin{equation}
    \delta X / X_{\rm no \, planet} \equiv (X - X_{\rm no \, planet}) / X_{\rm no \, planet} \, ,
    \label{eqn:dfn-perturbation}
\end{equation}
where $X_{\rm no \, planet}$ is the state of an identical ``no planet'' simulation\footnote{\citet{2020-sturm} also use an empty simulation as the unperturbed disk (private communication), but use notation $X_{0}$ instead of $X_{\rm no \, planet}$.
Normalizing with an empty simulation instead of the initial state is an easy way to account for radial drift-induced perturbations in the dust.}, at the same time snapshot. Our definition of perturbation is motivated from a theoretical perspective as it isolates the effect of the planet. Contrast is defined relative to the azimuthal average of the perturbed disk as
\begin{equation}
    {\rm contrast\, \,  in\, \, }X \equiv (X - \overline{X}_{\phi}) / \overline{X}_{\phi} \, ,
    \label{eqn:dfn-contrast}
\end{equation}
where $\overline{X}_{\phi}$ is the azimuthal average of the same disk, again at the same time. Our definition of contrast is motivated from an observational perspective, as an observer only has one disk to work with.

The hydrodynamic models presented in Figs. \ref{fig:sec3_stokes-azioffset-alma} \& \ref{fig:sec3_Sig-T-Inu} of this section are our shorter (15 orbits) set, described at the end of \S\ref{subsec:hydrosims}, which we conducted with a $\Mp = 1.0 \, \Mth$ planet and without a planet; Fig. \ref{fig:sec3_spiral-arm-contrast} uses our radiative transfer models (1500 orbits).

%%%%%%%%%%%%%%%%%%%%%%%%%%%%%%%%%%%%%%%%%%%%%%%%%%%%%%%%%%%%%%%%%%%%%%%%%%%%%%%%
\subsection{Do dust spirals look different to gas spirals?} \label{subsec:dustgas-coupling}
%%%%%%%%%%%%%%%%%%%%%%%%%%%%%%%%%%%%%%%%%%%%%%%%%%%%%%%%%%%%%%%%%%%%%%%%%%%%%%%%

\begin{figure*}
\plotone{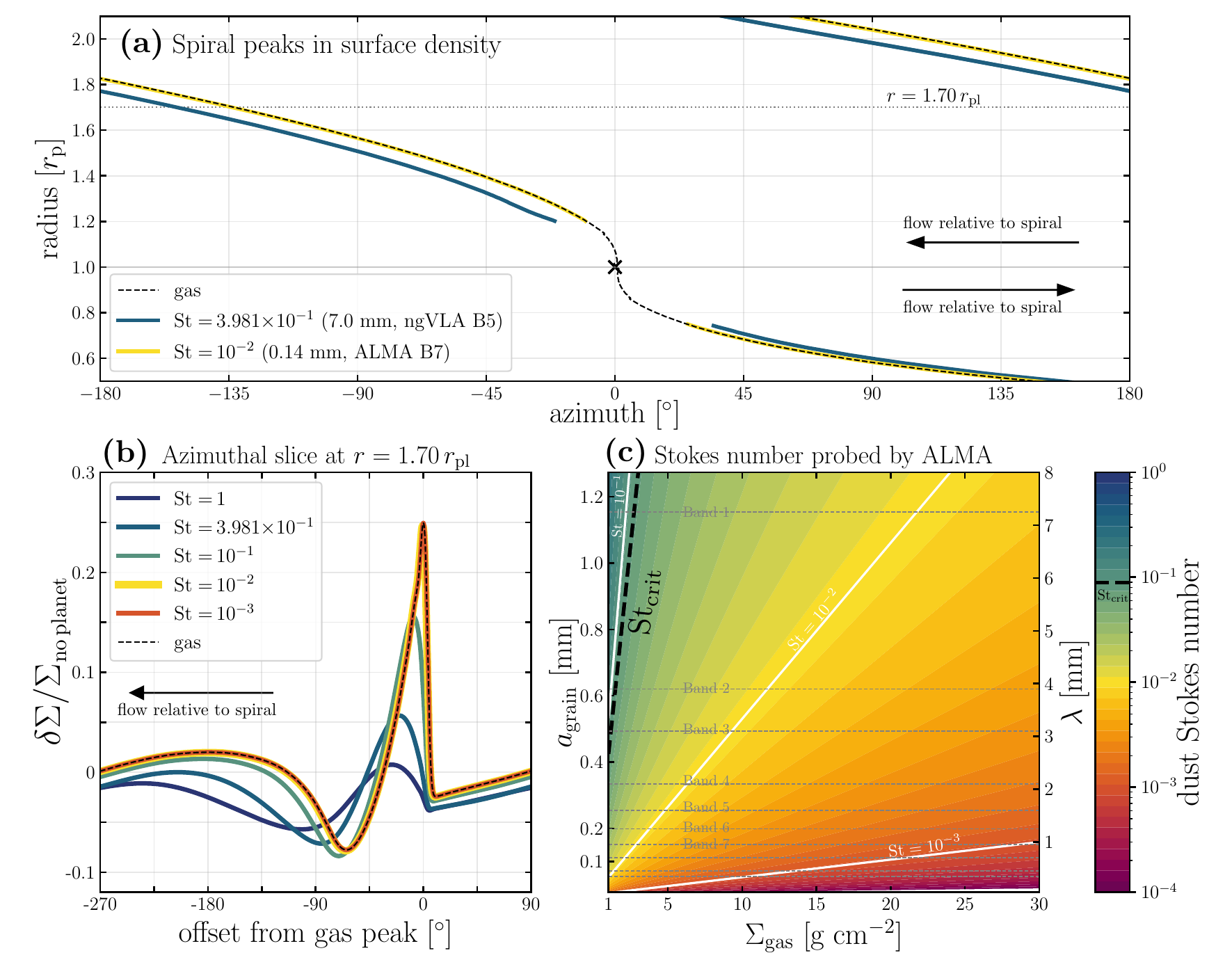}
\caption{Spiral morphology in surface density, and expectations for ALMA observations. \textbf{(a)} The spine of the inner and outer primary spiral arm in the gas, one species of well-coupled dust ($\St = 10^{-2}$, or $\agrain = 0.14\, \mm$ if $\Sigg=3\,\gcmsqrd$) and one of poorly coupled dust ($\St = 0.4$, $7.0\, \mm$), driven by a $1.0 \, \Mth$ planet in an adiabatic ($\beta=10$) disk. Poorly coupled dust ($\St > \Stcrit = 0.089$, Eqn. \ref{eqn:Stcrit}) forms spirals whose peaks azimuthally lag behind the gas. \textbf{(b)} An azimuthal slice of the surface density perturbation far from the planet ($r=1.7 \, \rp$), for gas and 5 species of dust (two well-coupled, one marginal, two poorly-coupled). In addition to being azimuthally offset, $\St > \Stcrit$ dust spirals have smaller amplitude. Note the yellow $\St = 10^{-2}$ curve has been made 2pts thicker to be visible behind the red $\St = 10^{-3}$ curve. \textbf{(c)} Stokes number calculated for a range of gas surface densities and ALMA dust grain sizes. For typical observing wavelengths of each ALMA band, we mark the dust grain size probed assuming (LH y-axis) $\agrain = \lambdaobs / 2\pi$ (RH y-axis). In general, ALMA probes well-coupled dust spirals. 
\label{fig:sec3_stokes-azioffset-alma}}
\end{figure*}

When a dust particle encounters the spiral wave, it experiences a temporary additional drag force due to the gas velocity perturbation and is disturbed from its near-Keplerian orbit. The degree of dust-gas coupling 
will determine the morphology of the resulting dust spiral. 
Throughout this paper, we use the term ``well-coupled'' to describe dust that forms spirals morphologically identical their gas spiral counterparts, and ``poorly-coupled'' to refer to dust whose spirals are morphologically different.
In this section we investigate the morphology of well-coupled and poorly-coupled dust spirals, and pinpoint the Stokes number that divides the two regimes.

\textit{Location of spiral peaks:} In Figure \ref{fig:sec3_stokes-azioffset-alma}(a), we trace the peaks of the spiral surface density perturbations (Eqn. \ref{eqn:dfn-perturbation}) in the gas and two species of dust, $\St = 4 \times 10^{-1}$ and $\St = 10^{-2}$. In comparing them to the gas, we see that these two species represent examples of poorly-coupled and well-coupled dust,  
respectively. The spine 
of the well-coupled dust spiral overlaps with that of the gas spiral perfectly throughout the disk. On the other hand, the poorly-coupled dust spiral lags behind the gas; the azimuthal offset between the two at $r=1.7 \, \rp$ for example is 20.04 degrees. 

\textit{Spiral amplitude:} In Figure \ref{fig:sec3_stokes-azioffset-alma}(b), we show an azimuthal cross section of the spiral surface density perturbation in the gas and five species of dust -- two Stokes numbers representative of well-coupled dust ($\St = 10^{-3}$, $ 10^{-2}$), one marginal ($\St = 10^{-1}$), and two poorly-coupled ($\St = 4\times 10^{-1}$, $1$). 
In addition to lagging behind the gas, the poorly-coupled dust spiral peaks are lower in amplitude. In contrast, 
the well-coupled dust spirals are indistinguishable from the gas. 
The marginal case suggests the de-coupling boundary occurs between $\St=10^{-2}$ and $\St=10^{-1}$.

\textit{What is ``well-coupled'' dust in the context of planet-driven spirals?} We can estimate the Stokes number that divides the well-coupled and poorly coupled regimes with a timescale argument, first described by \citet{2020-sturm}, where we compare the time it takes a dust particle to cross the gas spiral wake to the particle's intrinsic stopping time. First, we define $\deltaphispiral$ to be the azimuthal width of the gas spiral as a fraction of a full revolution, measuring it 
as the full width at half maximum (FWHM)\footnote{In contrast to \citet{2020-sturm}, who estimate it using a Gaussian fit.} in the perturbed surface density (i.e. the ``foot'' is at $\delta \Sigma /\Sigemp=0$ in the gas).
Note that $\deltaphispiral$ is a function of, at a minimum, planet mass and radius. With that number, we calculate the gas spiral crossing time \cite[Eqn. 5 of][]{2020-sturm}:
\begin{equation}
    \tcross = \deltaphispiral \, {\tdyn \over (1 - \Omp \, \tdyn)} \, ,
    \label{eqn:tcross}
\end{equation}
where $\tdyn = 1/\Omk$ is the dynamical time and $\Omp$ is the planet's (and therefore the gas spiral's) angular velocity. This equation assumes the dust moves at Keplerian velocity, 
and takes into account the additional time a particle spends in the spiral because the spiral moves in the same direction as the Keplerian flow.

Following \citet{2020-sturm} we define the ``critical'' Stokes number --the boundary between what constitutes well-coupled and poorly-coupled dust-- as the Stokes number for which $\tcross$ and $\tstop$ (Eqn. \ref{eqn:stokes-number}) are equal:
\begin{equation}
    \Stcrit = {\deltaphispiral \over (1 - \Omp \, \tdyn)} \, .
    \label{eqn:Stcrit}
\end{equation}
For the disk and planet parameters of Figure \ref{fig:sec3_stokes-azioffset-alma}, we find $\deltaphispiral= 0.05$ and $\Stcrit= 0.089$ at $r=1.7 \rp$, in agreement with the results presented in the top and bottom left panels. The critical Stokes number
offers an intuitive picture: Dust with Stokes number higher than $\Stcrit$ takes longer to respond to the gas spiral drag forces than the amount of time those forces act on them, and so their spiral morphology is different.

\textit{What are the implications for ALMA observations?}
In Figure \ref{fig:sec3_stokes-azioffset-alma}(c), we calculate the dust Stokes number
for a range of gas surface densities $\Sigma_{\rm gas}$ and dust sizes $\agrain$ (Eqn. \ref{eqn:stokes-number}). The horizontal lines mark the ALMA bands at which dust of a certain size is probed the best, assuming the median observing wavelength at each band $\lambdaobs=2\pi\agrain$ (Eqn. \ref{eqn:agrain-lambdaobs}). 
$\Stcrit = 0.089$ (dashed black line)
delineates the well-coupled and poorly-coupled regimes. This panel shows that, at almost all observing wavelengths and gas surface densities, ALMA probes well-coupled dust ($\St < \Stcrit$). We therefore expect that planet-driven dust spirals observed by ALMA 
will be 
perfect tracers of their parent gas spirals at the disk midplane.

%%%%%%%%%%%%%%%%%%%%%%%%%%%%%%%%%%%%%%%%%%%%%%%%%%%%%%%%%%%%%%%%%%%%%%%%%%%%%%%%
\subsection{The Ingredients of Intensity: In what disks are dust spirals most prominent?} \label{subsec:temperature-spirals}
%%%%%%%%%%%%%%%%%%%%%%%%%%%%%%%%%%%%%%%%%%%%%%%%%%%%%%%%%%%%%%%%%%%%%%%%%%%%%%%%

\begin{figure*}
\plotone{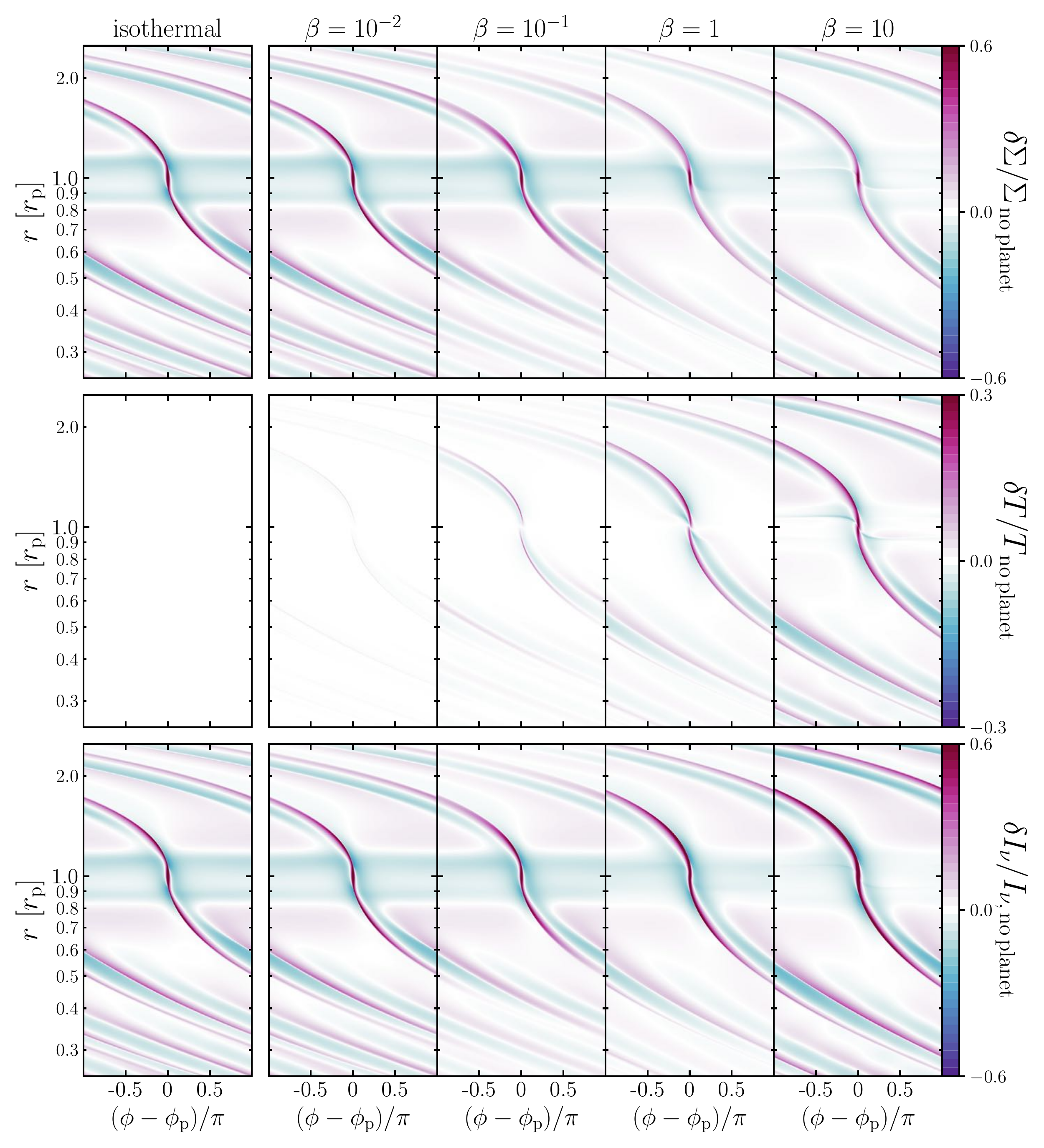}
\caption{Planet-induced spiral perturbations (Eqn. \ref{eqn:dfn-perturbation}) in
disks with increasing cooling timescales $\beta$ (\textbf{left to right}).
The spiral dust surface density perturbation (\textbf{top}) 
is largest in isothermal disks, whereas the temperature perturbation (\textbf{middle}) is necessarily zero in isothermal disks and increases monotonically with the cooling time.
As a result, the observable dust intensity perturbation (\textbf{bottom}),
shown here in the optically thin limit ($I_\nu \propto \Sigd \, B_\nu(\Td)$, where $\nu = 345 \, \GHz$), is largest in adiabatic disks that cool slowly. See also Fig. \ref{fig:app_multibeta} in \S\ref{sec:app:sec3-supplementary}. The planet mass is $\Mp = 1.0 \, \Mth$. Note that the colourbar range in the temperature row is half the value in the other two rows. 
\label{fig:sec3_Sig-T-Inu}}
\end{figure*}

Both the surface density of the dust and its temperature contribute to the overall dust thermal emission that we detect with ALMA continuum observations. It is therefore relevant for the observability of planet-driven dust spirals to understand how the spirals manifest in both the dust surface density and temperature, and to disentangle each ingredient's contribution to the surface brightness of the spiral.

From top to bottom, Figure \ref{fig:sec3_Sig-T-Inu} shows the perturbations (Eqn. \ref{eqn:dfn-perturbation}) in dust surface density, dust temperature, and the resultant dust intensity in the optically thin limit generated by a $1.0 \, \Mth$ planet. As we are interested in ALMA observations, we focus on well-coupled dust ($\St \ll \Stcrit$; \S\ref{subsec:dustgas-coupling}).
From left to right, we show these quantities in a locally isothermal disk and in disks with an adiabatic equation of state and different dimensionless cooling timescales. The dependence of these three perturbation quantities on the cooling rate is not always monotonic, and the five selected cases ($\beta = 10^{-2}$, $10^{-1}$, $1$, $10$ and the isothermal EoS representing the limit $\beta \rightarrow 0$) represent the full range of behaviours for the thermodynamics we consider.
Complementing this figure is Fig. \ref{fig:app_multibeta} in \S\ref{sec:app:sec3-supplementary}. 

\textit{Dust surface density:} Consider first the surface density perturbation, shown in the top row of Figure \ref{fig:sec3_Sig-T-Inu}. 
In the entire domain, the density spiral amplitude is strongest in the locally isothermal disk, while it drops by $\lesssim 10\%$ when $\beta$ increases to $10^{-2}$.
At the other extreme, the density spiral amplitude in disks with 
$\beta \geq 10$ are $\sim50\%$ as strong as the isothermal case near the planet ($0.6 \, \rp < r < 2.0 \, \rp$) but become more comparable further away.
In the intermediate cases, $10^{-1} \leq \beta \leq 1$, the spiral density waves also start $\sim50\%$ as strong as the isothermal case near the planet, but are significantly damped as they propagate. These results agree with those presented in \citet[their Fig. 1]{2020a-miranda-rafikov} and \citet[their Fig. 9]{2020-zhang-zhu-cooling}; efficient cooling leads to stronger compression. 
If we considered the well-coupled dust surface density alone, we might naively expect spirals to have higher contrasts, and thus be easier to detect given the same background disk surface brightness,
in disks with short cooling timescales.

\textit{Dust temperature:} In the middle row of Figure \ref{fig:sec3_Sig-T-Inu}, we present the temperature perturbation.
It is the strongest near the planet and decreases as the waves propagate in both directions. Here, the dependence on the cooling timescale 
is monotonic
--- the temperature spiral amplitude at all radii is an increasing function of $\beta$,
as inefficient cooling results in larger temperature increase from adiabatic compression. The results are in agreement with \citet{2021-muley}.

\textit{Dust intensity:} The bottom row of Figure \ref{fig:sec3_Sig-T-Inu} shows the 
intensity perturbation in the optically thin limit,
i.e., $I_{345 \, {\rm GHz}} \propto \Sigd \, B_{345 \, {\rm GHz}}(\Td)$.
In this way we demonstrate the effect of the temperature spiral without the added complication of disk optical depth, as a start. The surface density and temperature perturbations combine to give the intensity spiral amplitude a non-monotonic dependence on the dimensionless cooling timescale.
It is largest in adiabatic disks with $\beta \geq 10$ at all disk radii. In other words, dust spirals are more prominent in slow cooling disks.
The inner and outer primary spiral amplitude in the $\beta = 10$ case is uniformly $\sim 20\%$ higher than in the locally isothermal disk (Fig. \ref{fig:app_multibeta} in \S\ref{sec:app:sec3-supplementary}).

A second crucial effect of the temperature spiral --or more fundamentally, the dimensionless cooling timescale-- is to introduce a degeneracy between the intensity spiral amplitude and the planet mass. Comparing the two most different cases, the intensity perturbation amplitude at $r= 1.7 \, \rp$ is 0.13 for $\beta = 10^{-1}$ and 0.45 for $\beta = 10$. This is a factor of 3.5 different for the same $1.0 \, \Mth$ planet.\footnote{Fixing instead the cooling timescale to $\beta=10$ and varying the planet mass by a factor of $10$ (from $0.3 \, \Mth$ to $3.0 \, \Mth$), the amplitude difference is a factor of $2.4$.}

\begin{figure*}
\plotone{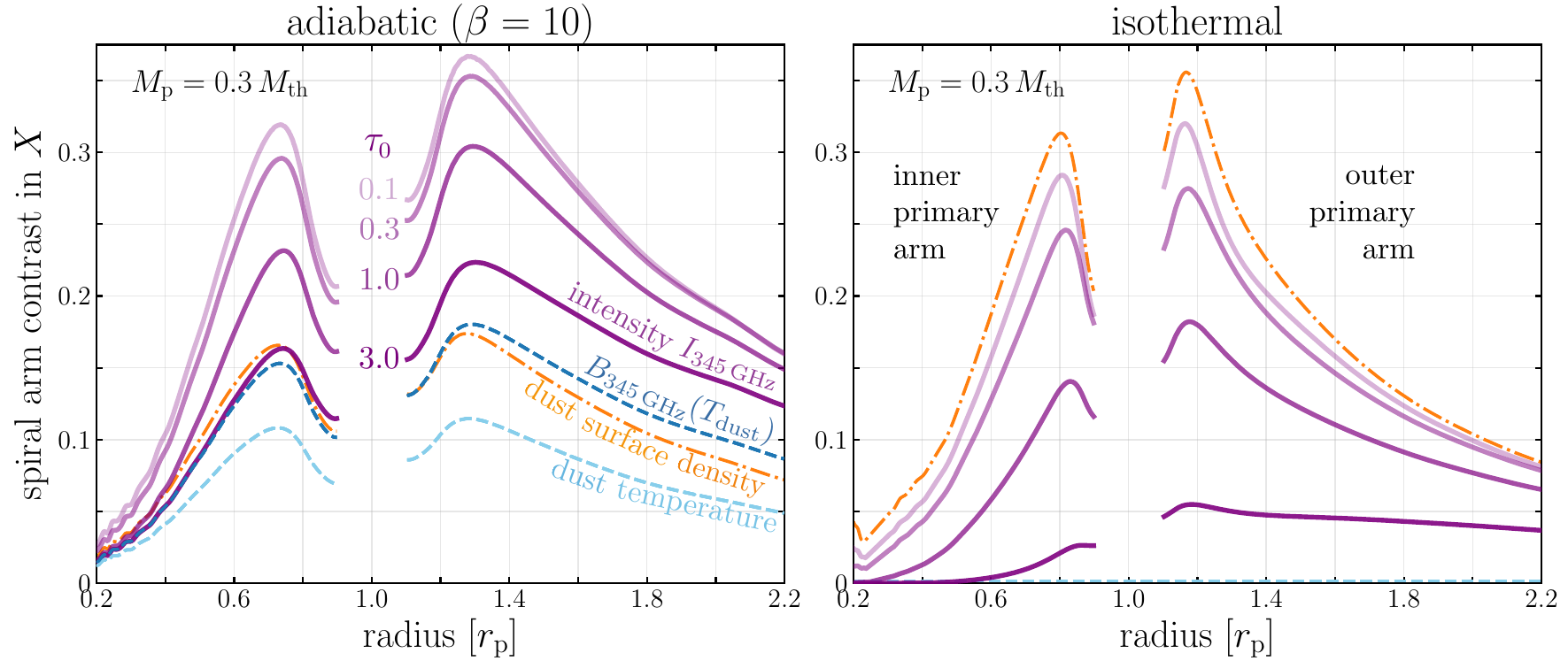}
\caption{Spiral arm contrast (Eqn. \ref{eqn:dfn-contrast}) 
in dust surface density (\textbf{orange dot-dashed}), temperature (\textbf{light blue dashed}), $B_{\rm 345 \, GHz}(\Td)$ (\textbf{dark blue dashed}) and intensity (\textbf{shades of solid purple}) traced along the inner and outer primary arms induced by a $0.3 \, \Mth$ planet. The shades of purple distinguish the dust intensity at different disk optical depths, from light (optically thin) to dark (optically thick); $\tauz$ is the initial optical depth at $\rp$. The dust surface density spiral contrast
in the isothermal case (\textbf{right}) is higher than that in the adiabatic one (\textbf{left}), but the former represents the upper bound of an isothermal spiral's contrast in intensity. Due to the presence of a temperature spiral in the adiabatic disks, the adiabatic spiral intensity contrast can exceed its dust surface density contrast and be brighter than in isothermal disks -- particularly at high $\tau$.
\label{fig:sec3_spiral-arm-contrast}}
\end{figure*}

In Figure \ref{fig:sec3_spiral-arm-contrast} we bring back the complication of disk optical depth by using the results of our radiative transfer calculations,
and switch to our observationally-motivated metric to quantify the spiral, the ``contrast'' (Eqn. \ref{eqn:dfn-contrast}). 
Figure \ref{fig:sec3_spiral-arm-contrast} shows the contrast traced along the inner and outer primary spiral arms in dust surface density, temperature and intensity, where intensity has now been calculated with the radiative transfer equation (Eqn. \ref{eqn:radiative-transfer}) for a range of disk optical depths. As per the first factor of Eqn. \ref{eqn:radiative-transfer}, we also plot the contrast in $B_{345 \, \GHz}(\Td)$. The driving planet has a mass $\Mp = 0.3 \, \Mth$. We show the results for the locally isothermal disk and compare them to a disk with an adiabatic EoS and 
$\beta = 10$.

Comparing the spiral surface density contrast between the two disks first, we see again that the isothermal case exhibits the most prominent density spirals. The difference is most significant (roughly a factor of 2) close to the planet (within 
$0.4 \,r_{\rm p}$ on either side). However, the spiral density contrast in the isothermal case constitutes an upper bound for how \textit{bright} spirals can be in that disk. For any non-zero optical depth, the isothermal spiral intensity contrast is lower than its dust density contrast.

In adiabatic disks on the other hand, the spiral intensity contrast can be larger than the density contrast, depending on the significance of the temperature spiral. In the $\beta = 10$ case, the spiral temperature contrast is $\sim 70 \%$ as strong as the density contrast, and is enhanced by the Planck function to give a contrast in $B_{345 \, \GHz}(\Td)$ that is $\sim 110\%$ as strong. This results in a spiral intensity contrast that not only exceeds its own density contrast, but that is also larger than the intensity contrast in the locally isothermal disk.  

This difference between spirals in isothermal and adiabatic disks becomes more pronounced when the disk is optically thick. For an initial optical depth at the planet's orbital radius of $\tauz = 3.0$, the spiral intensity contrast in the adiabatic $\beta = 10$ disk is 3-4 times larger than in the isothermal disk. The temperature spiral in adiabatic disks represents a ``floor'' for the spiral intensity; as $\tau$ increases it takes on a greater fraction of the responsibility for making adiabatic spirals brighter than isothermal ones, while isothermal spiral contrast disappears when $\tau \rightarrow \infty$.

Finally, we note that the outer spiral arm can have $10-40$\% larger contrast in intensity than the inner arm at its peak at various optical depths. This, combined with that the outer spiral fades more slowly as it propagates away from the planet, makes the outer spiral easier to identify in observations.

%%%%%%%%%%%%%%%%%%%%%%%%%%%%%%%%%%%%%%%%%%%%%%%%%%%%%%%%%%%%%%%%%%%%%%%%%%%%%%%%
%%%%%%%%%%%%%%%%%%%%%%%%%%%%%%%%%%%%%%%%%%%%%%%%%%%%%%%%%%%%%%%%%%%%%%%%%%%%%%%%
\section{Synthetic ALMA Observations: An Observer's Guide} \label{sec:results2}
%%%%%%%%%%%%%%%%%%%%%%%%%%%%%%%%%%%%%%%%%%%%%%%%%%%%%%%%%%%%%%%%%%%%%%%%%%%%%%%%
%%%%%%%%%%%%%%%%%%%%%%%%%%%%%%%%%%%%%%%%%%%%%%%%%%%%%%%%%%%%%%%%%%%%%%%%%%%%%%%%

\begin{figure*}
\plotone{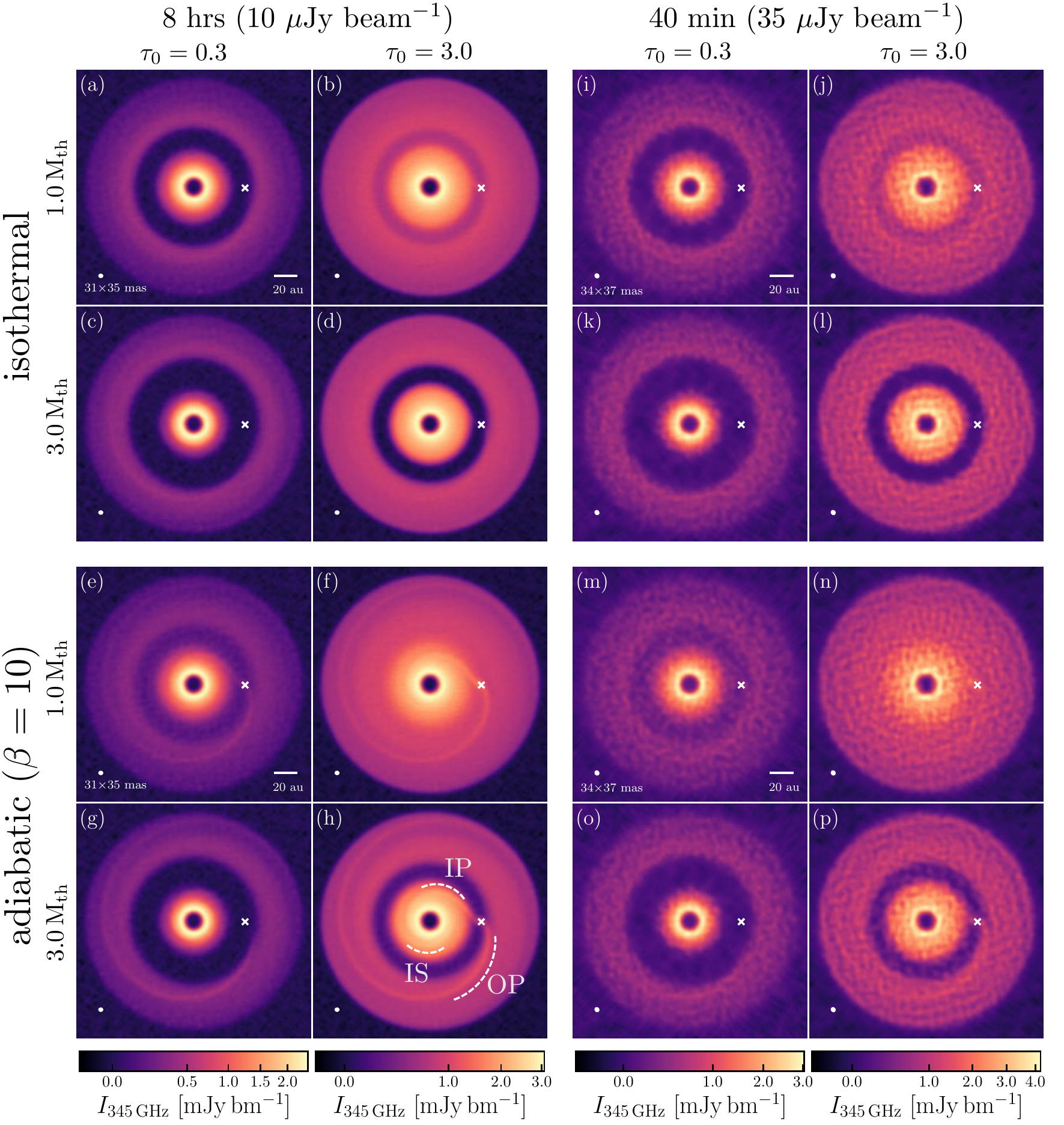}
\caption{A selection of synthetic ALMA B7 continuum images with the C43-5 + C43-8 configuration pair, demonstrating trends with equation of state (\textbf{top} vs. \textbf{bottom}), combined on-source time (\textbf{left} vs. \textbf{right}), planet mass (\textbf{inner top} vs. \textbf{inner bottom}), and disk optical depth (\textbf{inner left} vs. \textbf{inner right}). The requested sensitivity (number in brackets, e.g. $10\, \muJybeam$) is what was used to determine the combined on-source time (e.g. 8 hrs).
The synthesized beam is shown in the bottom left corner of each image, and each colourbar applies to the whole column. Images are shown with a ${1 \over 2}$-powerlaw stretch. The full set of images is available at \href{https://doi.org/10.6084/m9.figshare.19148912}{https://doi.org/10.6084/m9.figshare.19148912}.
\label{fig:sec4_spread-Inu}}
\end{figure*}

\begin{figure*}
\plotone{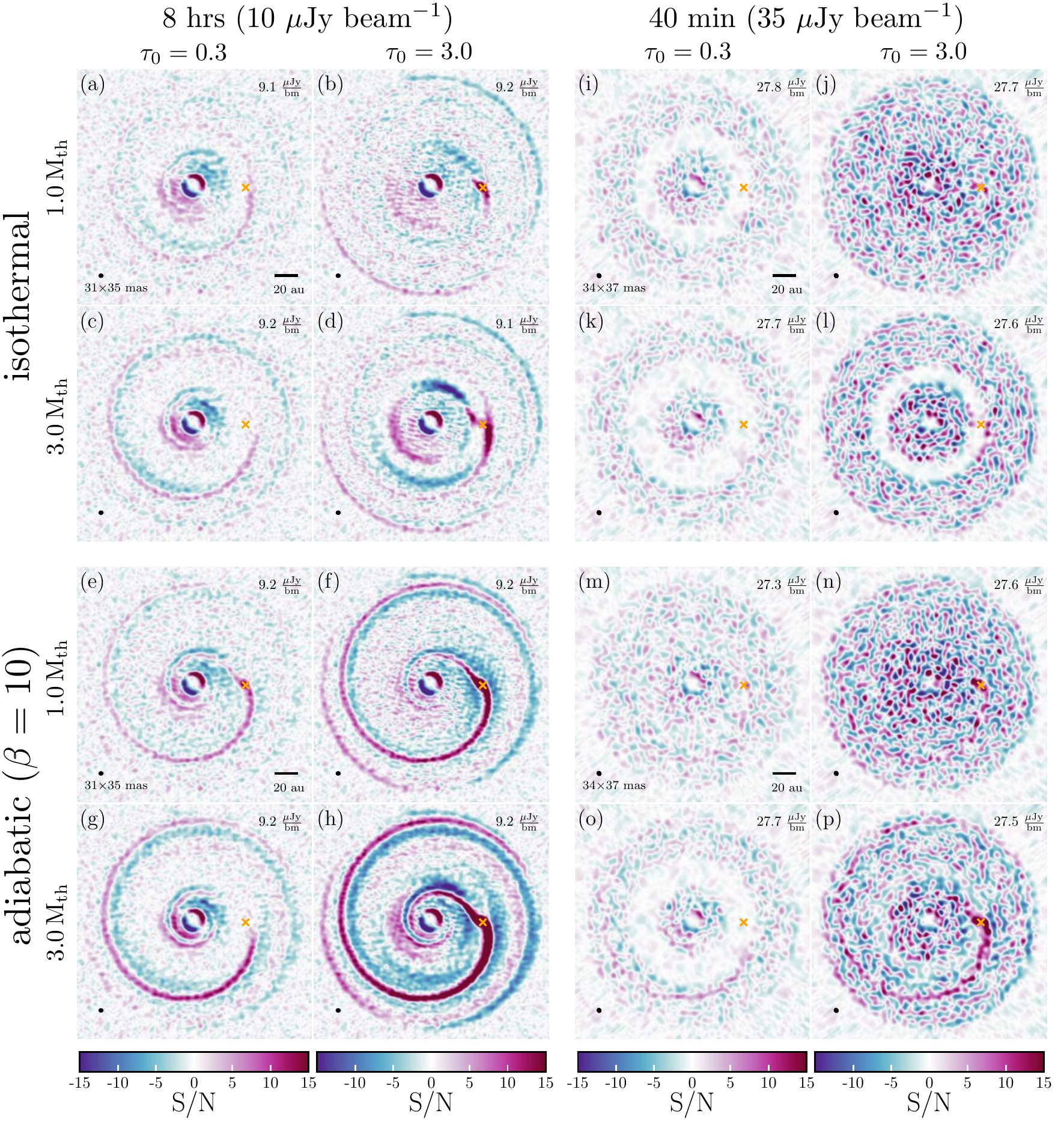}
\caption{Residual maps of the observations in Figure \ref{fig:sec4_spread-Inu}. The signal to noise ratio is calculated as ${\rm S/N} = (I_{\rm obs} - \overline{I_{\rm obs}}_{\phi})/({\rm rms\, noise})$, where $\overline{I_{\rm obs}}_{\phi}$ is an axisymmetric disk obtained by azimuthally averaging in the image plane. The measured rms noise in each observation after cleaning is written in the top right corner of each panel, and in all cases is $\sim 10-20\%$ less than the requested sensitivity.
\label{fig:sec4_spread-SN}}
\end{figure*}

In this section, we present synthetic ALMA B7 continuum observations and our method of highlighting the planet-driven spiral signal. We report trends in the amount of ALMA time needed under different disk and observing conditions (\S\ref{subsec:alma-time}); we describe the impact of different ALMA antenna configurations (\S\ref{subsec:angular-resolution}); we feature a successful recovery of the spiral driven by a low mass planet (\S\ref{subsec:lowmass-planets}); and we show the impact that gaps and rings could have on our ability to recognize co-located spiral arms (\S\ref{subsec:gaps-rings}).

In Figure \ref{fig:sec4_spread-Inu} we show a gallery of 16 synthetic ALMA continuum images, a representative selection from our full set\footnote{The results from our full set ($3$ planet masses, $\times 4$ disk optical thicknesses, $\times 2$ equations of state, $\times 3$ antenna configuration pairs, $\times 6$ integration times $=432$ model images and their residual maps) plus an additional set of inclined disk models are available at \href{https://doi.org/10.6084/m9.figshare.19148912}{https://doi.org/10.6084/m9.figshare.19148912}.}, obtained with the C43-5 + C43-8 configuration pair (beam size $31\times35 \, \mas$). This gallery demonstrates the outcomes under a variety of disk and observing conditions: two integration times (8 hours vs. 40 mins on-source), two equations of state (adiabatic with a cooling timescale $\beta=10$ vs. isothermal), two disk optical depths (marginally optically thin $\tauz=0.3$ vs. optically thick $\tauz=3.0$) and two planet masses ($\Mp = 1.0 \, \Mth$ or $3.0 \, \Mth$). In panels (e), (f), (g), (h) and (p),  
the outer primary spiral arm is very clearly visible in the continuum image. In panels (f) and (h), the inner primary (and panel (h), the inner secondary) arm can also be seen. We have labelled the outer primary (``OP''), inner primary (``IP'') and inner secondary (``IS'') arms in panel (h).

In order to quantify the robustness of these detections, and in order to amplify the spiral signal in less conspicuous cases, we make residual maps for all our synthetic observations
-- in the image plane. We first transform the ALMA image in on-sky coordinates into polar ($r$,$\phi$) coordinates\footnote{
We use the \texttt{polarTransform} python module: https://polartransform.readthedocs.io/en/latest/}, then average the observed intensity over the full azimuth to obtain an axisymmetric disk, %$I_{\phi\rm\ averaged}$
$\overline{I_{\rm obs}}_{\phi}$. 
We subtract that map from the observed image and normalize the difference by the observation's rms noise, such that the reported quantity in 
residual maps is ${\rm S/N} = (I_{\rm obs} - \overline{I_{\rm obs}}_{\phi})/({\rm rms\, noise})$. Figure \ref{fig:sec4_spread-SN} presents such residual maps for the observations in Figure \ref{fig:sec4_spread-Inu}. 

The strongest 
dust spiral recovery in all our permutations of disk conditions is the one in panel (h) of Figures \ref{fig:sec4_spread-Inu} \& \ref{fig:sec4_spread-SN}: a $3.0 \, \Mth$ planet in an adiabatic ($\beta=10$), optically thick ($\tauz = 3.0$) disk. Change the disk's equation of state to locally isothermal [panel (d)], and the spiral hardly appears in the residuals at all. Decreasing the planet mass to $1.0 \, \Mth$ [panel (f)] decreases the spiral signal in ${\rm S/N}$, but it is still 
strong. Decrease the disk optical depth by a factor of 10 [panel (g)] and some parts of the inner primary arm are lost, but three spiral arms (outer primary, inner primary, inner secondary) are still visible in the residuals. The outer spiral driven by the planet in panel (h) is so significant that it is 
identifiable in the residuals of a 40 min observation [panel (p)].

From Figure \ref{fig:sec4_spread-Inu} we conclude that the signal from planet-driven dust spirals in isothermal disks is very weak in comparison to those in adiabatic disks that cool slowly. We also see that even though spiral intensity 
contrasts are intrinsically larger in optically thin disks (Fig. \ref{fig:sec3_spiral-arm-contrast}), they are actually more difficult to observe due to the disk being dimmer overall.

Before moving on to permutations of observing conditions, a few remarks on residual maps. Firstly, we find that they are an excellent tool for probing spirals, though this comes with the caveat that they are easy to make for our face-on model disks
and may not be as straightforward to make for real disk observations. 
Secondly, we emphasize the importance of plotting both the positive \textit{and negative} residuals when searching for planet-driven spirals. Detecting both a spiral's peak and its associated trough can lend weight to the detection of the structure, as well as to the conclusion that it is companion-driven, because they are predicted together by density wave theory \cite[e.g.,][]{2018a-bae-zhu}. We also tried making residual maps in the visibility plane using \texttt{frankenstein} \citep{2020-jennings-frank}, which has been shown to produce accurate fits to real visibility data \cite[e.g.,][]{2021-jennings-dsharp}, but with our models we found that the imaged \texttt{frank} visibility residuals showed the spiral less clearly (see Fig. \ref{fig:app_frank} in \S\ref{sec:app:sec4-supplementary}).

%%%%%%%%%%%%%%%%%%%%%%%%%%%%%%%%%%%%%%%%%%%%%%%%%%%%%%%%%%%%%%%%%%%%%%%%%%%%%%%%
\subsection{How much ALMA time do you need?} \label{subsec:alma-time}
%%%%%%%%%%%%%%%%%%%%%%%%%%%%%%%%%%%%%%%%%%%%%%%%%%%%%%%%%%%%%%%%%%%%%%%%%%%%%%%%

\begin{figure*}
\plotone{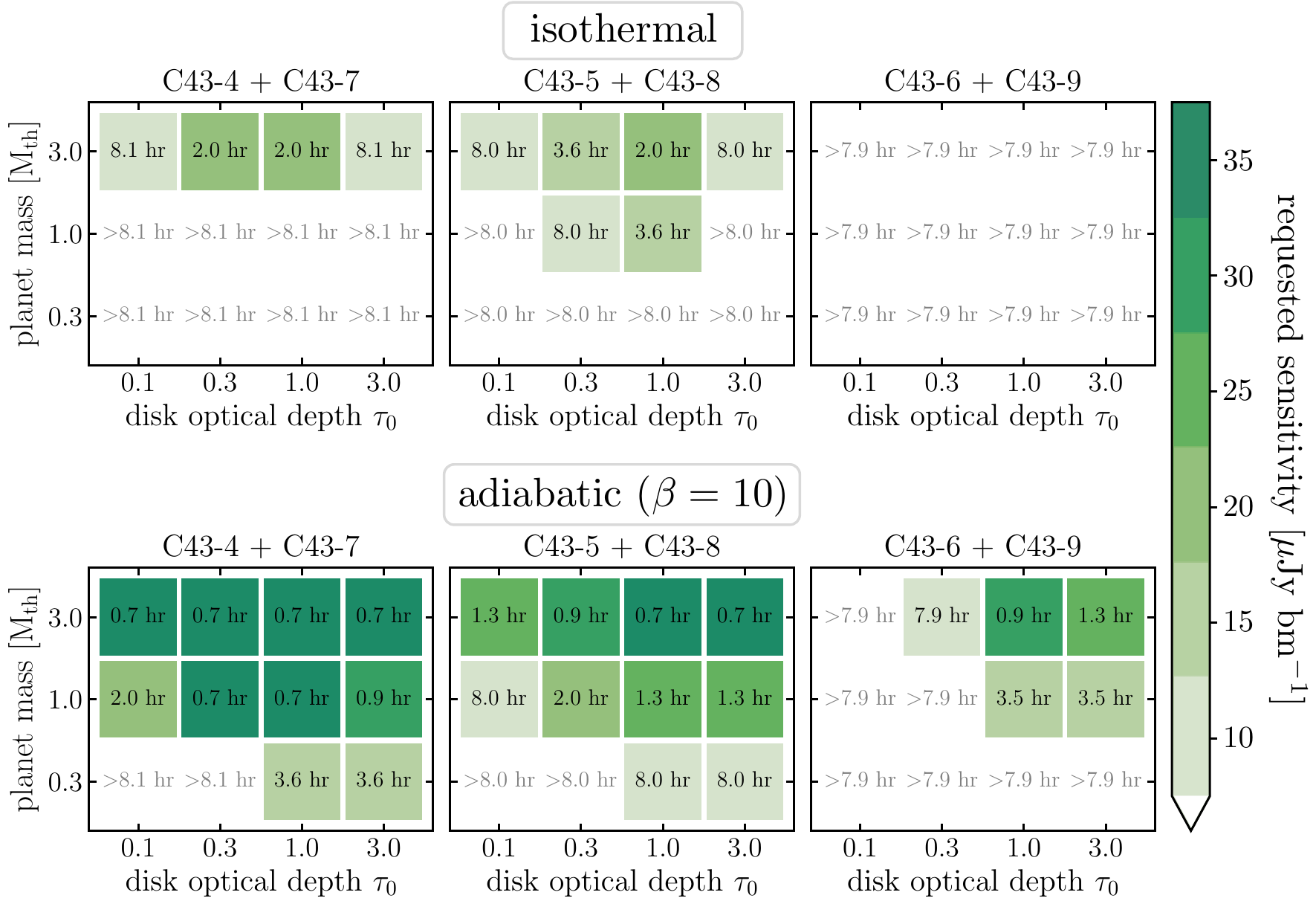}
\caption{Amount of on-source time required to recover the planet's outer spiral arm in residuals of our B7 ALMA continuum images under the disk and observing conditions we explore in this work, coloured by the corresponding requested sensitivity. White indicates cases where we were not able to recover the spiral. 
\label{fig:sec4_multidim-table}}
\end{figure*}

It is \textit{a priori} possible that observing planet-driven spirals requires very deep observations. To test this possibility, we created ALMA observations achieving requested sensitivities of 10, 15, 20, 25, 30, 35 $\muJybeam$ with 0.7, 0.9, 1.3, 2.0, 3.5 and 8.0 hours of combined on-source time (totaling roughly 2, 2.5, 3.8, 5.0, 9.0 and 20 hours with overheads) and judged whether the spiral was detected in each case. For context, the DSHARP program 
had a median integration time of $\sim 1.4$ hours \citep{2018-andrews-dsharp1}, and the longest integration done on a single disk to date (achieving a sensitivity and beam size comparable to our models, $25.7 \, \muJybeam$ and $\thetaAR= 24.6 \, \mas$ in B7) is 5.59 hours of on-source time toward HL Tau (2019.1.01051.S, currently in progress). 
Our maximum on-source time thus pushes the envelope by $\sim 2.5$ additional on-source hours.

To decide on a criterion for successful spiral recovery, we considered: \textit{How does one recognize a spiral?} By definition, a spiral is a structure that extends some range in azimuth $\Delta \phi$ for each increment in radius $\Delta r$. The more azimuth $\Delta \phi$ we notice the structure spanning, the more confident we are that it is a spiral and not a segment of a circle, or an azimuthal asymmetry. Motivated by that concept, we set the criterion for a successful spiral recovery to be that we can trace the outer spiral arm in our residual maps with ${\rm S/N} = 5$ contours over at least 90 degrees continuously in azimuth (not necessarily starting from the known planet location). We chose to focus on the outer primary spiral 
%only 
because it has a larger intrinsic contrast (Fig. \ref{fig:sec3_spiral-arm-contrast}) and was more often apparent in our residual maps (Fig. \ref{fig:sec4_spread-SN}) than the inner primary arm. Structures at larger radii also naturally have larger angular extent.

Figure \ref{fig:sec4_multidim-table} is a visual table depicting the amount of on-source time required to recover the outer spiral arm under our 72 permutations of disk and observing conditions ($\times 2$ EoS $\times 3 \, \Mp$ $\times 4 \, \tauz$ $\times 3$ configuration pairs). We created this figure by identifying, for each permutation, the observation with the least on-source time that still recovered the spiral. That minimum on-source time is coloured by the corresponding ALMA OT requested sensitivity. White space (with grey text, e.g. ``$> 8.1$ hr'') indicates that we were not able to recover the spiral under those disk and observing conditions with our maximum tried on-source time.

Figure \ref{fig:sec4_multidim-table} shows that, if they are present, planet-driven dust spirals are easier to observe (i.e., require less integration time to detect) in adiabatic disks that cool slowly ($\beta \gtrsim 10$), that are 
marginally but not too optically thick ($\tauz \gtrsim 1.0$), and that host massive planets ($\Mp \gtrsim 1.0 \, \Mth$). In such disks,
spirals can be detected with integration times on the order of hours.

%%%%%%%%%%%%%%%%%%%%%%%%%%%%%%%%%%%%%%%%%%%%%%%%%%%%%%%%%%%%%%%%%%%%%%%%%%%%%%%%
\subsection{What angular resolution do you need?} \label{subsec:angular-resolution}
%%%%%%%%%%%%%%%%%%%%%%%%%%%%%%%%%%%%%%%%%%%%%%%%%%%%%%%%%%%%%%%%%%%%%%%%%%%%%%%%

\begin{figure*}
\plotone{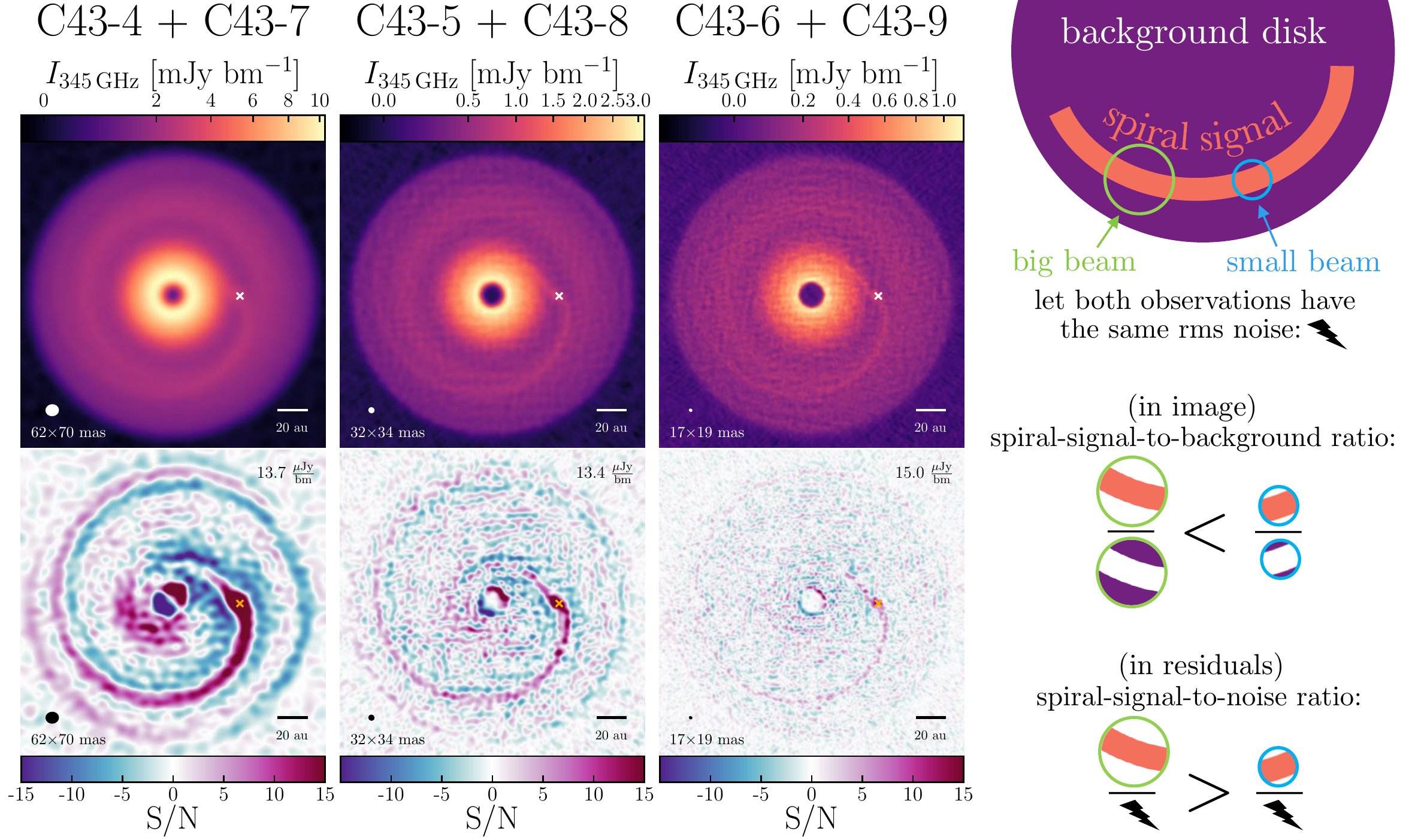}
\caption{Synthetic ALMA B7 continuum images (\textbf{top}) and residuals (\textbf{bottom}) observed with different configuration pairs, shown from left to right in order of low to high angular resolution, for a $1 \, \Mth$ planet embedded in an adiabatic ($\beta=10$), marginally optically thick disk ($\tauz=1.0$). In all observations, the measured rms noise (top right corner of residual panels) is very similar. The requested sensitivity is $15 \, \muJybeam$, corresponding to a combined on-source time of 3.5 hrs. The schematic (\textbf{right}) shows that a lower angular resolution observation has a lower spiral-signal-to-background ratio (the spiral is ``washed out'') but actually has higher spiral-signal-to-noise ratio (detects the spiral with the highest ${\rm S/N}$).
\label{fig:sec4_configuration}}
\end{figure*}

Planet-driven spiral arms are intrinsically fine structures with widths comparable to scale height (e.g. Fig. \ref{fig:sec3_Sig-T-Inu}).
As such, it is natural to think that high angular resolution observations --observations that \textit{resolve} the spiral-- are necessary in order to detect them. 
We find that the situation is more nuanced, and that high resolution doesn't necessarily lead to best detectability.

In Figure \ref{fig:sec4_configuration}, we show ALMA continuum observations and residual maps obtained with three different antenna configuration pairs: C43-4 + C43-7 ($\thetaAR = 0.061 \arcsec$),
C43-5 + C43-8 ($\thetaAR = 0.028 \arcsec$),
and C43-6 + C43-9 ($\thetaAR = 0.017 \arcsec$), translating to $8.5 \, \au$, $3.9 \, \au$ and $1.7 \, \au$ at $d=140 \, \pc$, respectively.
We achieve a measured sensitivity of $\sim$13-15 $\Jybeam$ with $\sim$3.5 hrs of on-source time in all cases. 
The disk is adiabatic ($\beta=10$), marginally optically thick ($\tauz=1.0$) and contains a $1.0 \, \Mth$ planet. 
If we approximate the spatial width of the spiral as $H(\rp)=3.5 \, \au$, we see that the C43-7 configuration does not resolve the spiral, C43-8 is marginal, and C43-9 resolves the spiral with $\sim 2$ beams.

In the continuum images (top panels of Fig. \ref{fig:sec4_configuration}), we find the intuitive result: the spiral signal is ``washed out'' in the lower angular resolution observation (left panel). In the higher angular resolution image (right panel), the outer spiral is easily seen directly in the image, and even the inner spiral arm is visible. The explanation is simply that what is captured by a smaller beam can be comprised of a greater proportion of spiral signal than background disk signal, in comparison to the proportion captured by a larger beam. In other words, when the spiral is not fully resolved,
a higher angular resolution observation has a higher spiral-signal-to-background ratio.

The nuance is introduced when one considers the \textit{robustness} of the spiral detection, i.e., the spiral-signal-to-\textit{noise} ratio. In the residual maps (bottom panels of Fig. \ref{fig:sec4_configuration}), we find that the lower angular resolution observation with C43-4 + C43-7 detects the spiral with the highest ${\rm S/N}$: we can trace ${\rm S/N}=5$ contours over $\sim270$ degrees, and ${\rm S/N}=10$ contours over $\sim 90$ degrees. On the other hand, the detection with the higher angular resolution observation (C43-6 + C43-9) is less robust; it did satisfy our recovery criterion (Fig. \ref{fig:sec4_multidim-table}), but only just.

The explanation for this rests on two factors: (1) the observed intensity is in units of $\Jybeam$, not $\Jyarcsec$; and (2) the observations we are comparing have very similar levels of rms noise. 
The spiral signal in $\Jybeam$ is larger for a larger beam (due to the beam's larger ``area''). 
Therefore, if noise is independent of beam size, the low angular resolution detection has a higher spiral-signal-to-noise ratio. This holds at all sensitivities that we explored. Of course, there is a minimum angular resolution required for spiral detection; one would need to resolve the distance between the spirals (think: Rayleigh criterion), and ideally any background structures on scales larger than the spiral.

The utility of archival or future continuum observations of real disks done at high angular resolution could be enhanced by uv-tapering to produce larger beam sizes, possibly improving the ${\rm S/N}$ and helping to identify the spiral signal in ${\rm S/N}$ space.

%%%%%%%%%%%%%%%%%%%%%%%%%%%%%%%%%%%%%%%%%%%%%%%%%%%%%%%%%%%%%%%%%%%%%%%%%%%%%%%%
\subsection{Detecting spirals from low mass planets?} \label{subsec:lowmass-planets}
%%%%%%%%%%%%%%%%%%%%%%%%%%%%%%%%%%%%%%%%%%%%%%%%%%%%%%%%%%%%%%%%%%%%%%%%%%%%%%%%

\begin{figure}
\plotone{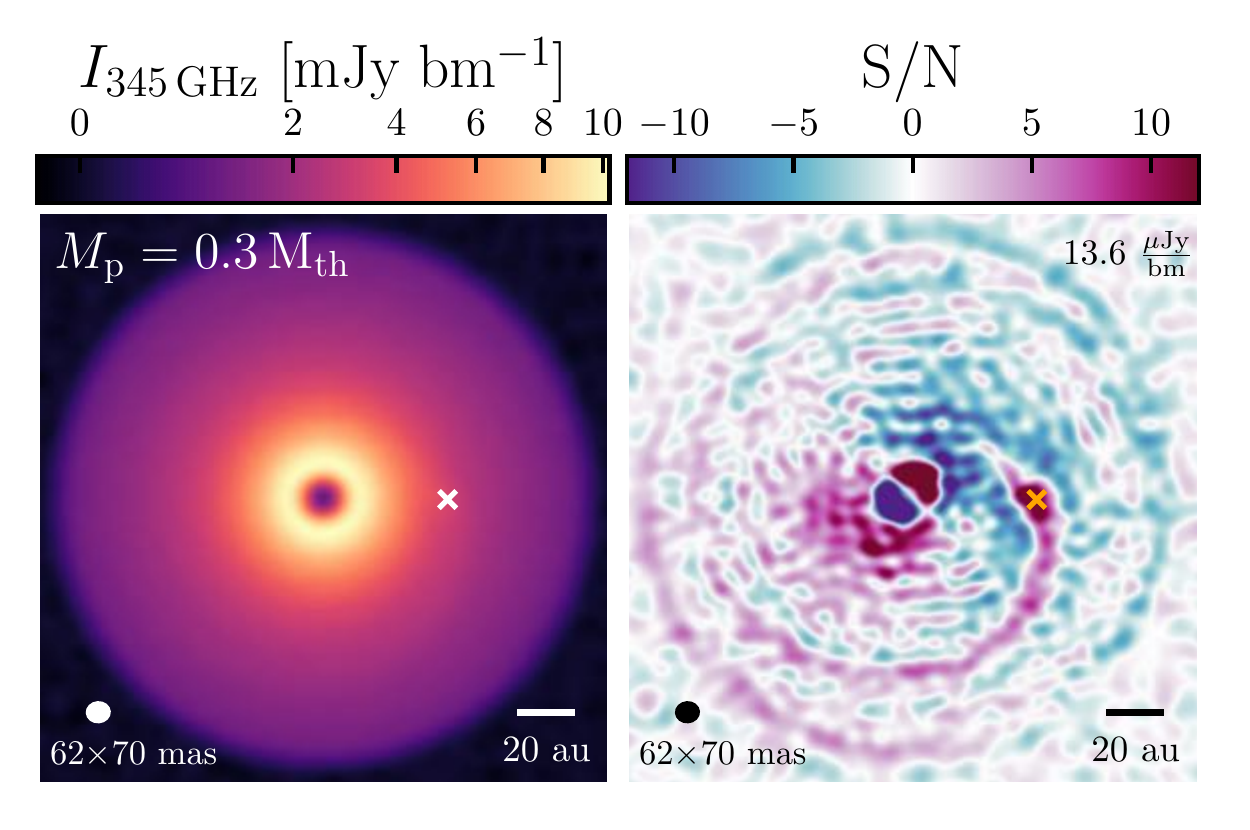}
\caption{Spiral arm recovery for a low mass planet ($\Mp = 0.3 \, \Mth =1.6 \, \Mnep$) in a marginally optically thick disk ($\tauz=1.0$) with an adiabatic equation of state ($\beta=10$), obtained with the combined C43-4 + C43-7 antenna configuration and a combined on-source time of 3.6 hours. The measured rms noise is $13.6 \, \muJybeam$.  
\label{fig:sec4_lowmass}}
\end{figure}

Detecting a low mass planet embedded in a protoplanetary disk is an exciting prospect because if we believe planets grow in mass over some non-zero formation timescale, then ``low mass'' translates to 
%``young'', 
``early stage'',
and probing lower mass planets probes closer to their birth. The challenge is linearly proportional to the reward, however, because the amplitude of planet-driven spirals (in the 
%low mass, linear 
sub-thermal mass regime) is proportional to the planet mass \citep{2011a-dong, 2018a-bae-zhu, 2019-miranda-rafikov}. We observe many dust gaps and rings, which we believe could be the birth-sites of planets, but as yet few co-located spirals. 
One possible explanation for the dearth of observed planet-driven spirals despite the abundance of observed gaps and rings could simply be that the spiral amplitude is too insignificant to be detected with ALMA. What is the lowest planet mass that drives a detectable dust spiral?

In Figure \ref{fig:sec4_lowmass}, we show a successful recovery of the outer spiral driven by our lowest mass planet, $0.3 \, \Mth$ ($0.1\, \Mjup$ or $1.6 \, \Mnep$, $q = 1.03 \times 10^{-4}$). As shown in Figure \ref{fig:sec4_multidim-table}, if the disk is adiabatic and cools slowly ($\beta = 10$), we can recover the outer spiral driven by this $0.3 \, \Mth$ planet in both the $\tauz = 1.0$ and $3.0$ disks after 3.6 hours of on-source time with the C43-4 + C43-7 configuration pair, and after 8.0 hours of on-source time with the C43-5 + C43-8 pair. Figure \ref{fig:sec4_lowmass} shows the 
former case with
$\tauz = 1.0$. The contrast (as we've defined contrast, Eqn. \ref{eqn:dfn-contrast}) of the outer spiral driven by the $0.3 \, \Mth$ planet in the adiabatic, $\tauz = 1.0$ disk ranges from 0.15 to 0.3 (see left panel of Fig. \ref{fig:sec3_spiral-arm-contrast}).

The spiral is not clearly visible directly in the synthetic ALMA continuum image, but is recovered in the residuals at ${\rm S/N} \geq 5$ over $\sim 160$ degrees of the disk, underscoring the utility of residual maps.
We caveat this successful recovery by noting that, although this planet had been living in our hydro simulations for 1500 orbits before we took its picture with ALMA, it created no observable gaps and rings, due to the 
%relatively high 
modest
viscosity adopted ($\alpha=10^{-3}$). As we discuss in the following section, the smoothness of the background disk aids the recovery.

%%%%%%%%%%%%%%%%%%%%%%%%%%%%%%%%%%%%%%%%%%%%%%%%%%%%%%%%%%%%%%%%%%%%%%%%%%%%%%%%
\subsection{Can 
%planet-driven 
spirals be hiding in gaps and rings?} \label{subsec:gaps-rings}
%%%%%%%%%%%%%%%%%%%%%%%%%%%%%%%%%%%%%%%%%%%%%%%%%%%%%%%%%%%%%%%%%%%%%%%%%%%%%%%%

\begin{figure*}
\plotone{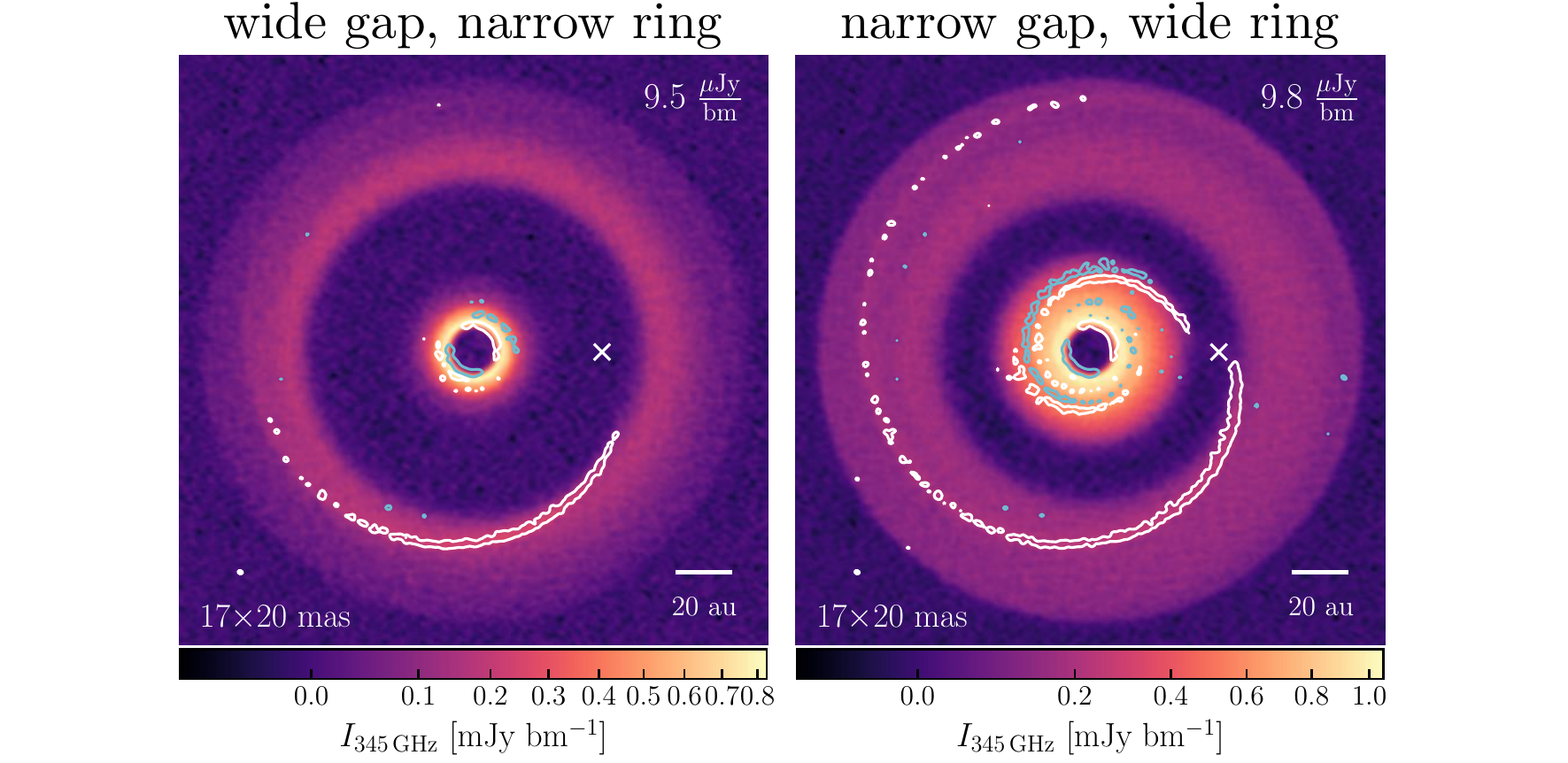}
\caption{Synthetic ALMA B7 continuum observations exploring the effect of dust gaps and rings on the spiral observability. Both disks are adiabatic ($\beta=10$) and contain a $3.0 \, \Mth \approx 1.0 \, \Mjup$ planet; both observations were obtained with the C43-6 + C43-9 configuration pair and 8.0 hours of on-source time. The overlaid white and blue contours are ${\rm S/N}=+5$ and  ${\rm S/N}=-5$ contours from the residual maps, respectively. The rms noise in each observation is written in the top right corner of the images and the beam size is marked at the lower left corner. Wide gaps and narrow outer rings reduce the amount of disk area over which the spiral can be traced.
\label{fig:sec4_gaps-rings}}
\end{figure*}

In this section, we explore another possible explanation for the dearth of observed planet-driven spirals despite the abundance of observed gaps and rings: Could the presence of the gaps and rings themselves be affecting the observability of the spirals? As discussed in \S\ref{subsec:alma-time}, an observer's ability to recognize a spiral as a spiral requires seeing it unfurl over a sufficiently large range of azimuth. To span more azimuth $\Delta \phi$, the spiral requires a larger radial breadth $\Delta r$ of smooth background disk -- and this can be affected by gaps and rings.

Figure \ref{fig:sec4_gaps-rings} compares the observability of spirals in a disk with a wide planet-induced gap and narrow outer ring (left panel), to that of a spiral in a disk with a narrower gap and wider outer ring (right panel). To compare the effect of the dust distribution alone, we have re-scaled the underlying intensity maps such that the outer ring in the two disks are equally bright. The left is our optically thinnest ($\tauz = 0.1$) disk model and the right is our marginally optically thick ($\tauz=1.0$) disk model, meaning their gas surface densities are different by a factor of 10; as a result the Stokes number of the $\agrain=0.14 \, \mm$ dust within them is also different by a factor of 10, giving the two disks their different dust distributions. Both disks have an adiabatic EoS ($\beta=10$) and embedded planet with mass $\Mp = 3.0 \, \Mth$. 
We show observations made with the C43-6 + C43-9 configuration pair and a combined on-source time of 8.0 hours (achieving measured rms noise $\sim 9.5 \, \muJybeam$) in order to investigate what might be considered a highly desirable observing scenario: high angular resolution and high sensitivity. Atop the observations we overlay ${\rm S/N} = \pm 5$ contours from the residual maps.% (no area in the residuals has been masked).

Comparing the two panels in Figure \ref{fig:sec4_gaps-rings}, we see that the presence of the wide gap and narrow ring renders a smaller fraction of the spiral visible in the contours. In the left panel, the outer spiral starts farther away from the planet and extends only $\sim 150$ degrees in azimuth due to the small available radial area. In the right panel, the spiral contour starts closer to the planet and extends a full $270$ degrees out to $\sim 2.2\, \rp$. Similarly, in the inner disk, there is very little recognizable evidence of a spiral in the left panel, whereas in the right panel we can trace the inner primary arm, the inner secondary arm, and the trough between them.

Like \citet{2019-miranda-rafikov-planet-interpretation}, we find that a given planet mass carves shallower gaps in disks with an adiabatic equation of state than in a locally isothermal disk (and this translates to the well-coupled dust distribution, e.g. compare panels (a) vs. (e), (b) vs. (f), and (d) vs. (h) in Fig. \ref{fig:sec4_spread-Inu}). This is another way that the disk equation of state can affect the observability of spiral arms, in addition to regulating their contrast. 

%%%%%%%%%%%%%%%%%%%%%%%%%%%%%%%%%%%%%%%%%%%%%%%%%%%%%%%%%%%%%%%%%%%%%%%%%%%%%%%%
%%%%%%%%%%%%%%%%%%%%%%%%%%%%%%%%%%%%%%%%%%%%%%%%%%%%%%%%%%%%%%%%%%%%%%%%%%%%%%%%
\section{Discussion} \label{sec:discussion}
%%%%%%%%%%%%%%%%%%%%%%%%%%%%%%%%%%%%%%%%%%%%%%%%%%%%%%%%%%%%%%%%%%%%%%%%%%%%%%%%
%%%%%%%%%%%%%%%%%%%%%%%%%%%%%%%%%%%%%%%%%%%%%%%%%%%%%%%%%%%%%%%%%%%%%%%%%%%%%%%%

\textbf{Inclined disks.} While most of our efforts were focused on face-on disks, we briefly experimented with inclined disks as well. Figure \ref{fig:app_inclined} in Appendix \S \ref{sec:app:inclined-disks} shows continuum images and deprojected residual maps for a demonstrative disk, inclined by $30^{\circ}$, $50^{\circ}$ and $70^{\circ}$, along with its original face-on model for comparison. We find that the spiral visibility in the deprojected residual maps is not significantly affected if the disk inclination is low ($\leq 50^{\circ}$), but becomes very unclear when the inclination is high ($\geq 70^{\circ}$). For a given target disk on the sky, the threshold inclination at which a spiral could no longer be as easily recovered in a deprojected residual map would depend on whether the beam resolves the structure along the disk's minor axis.

% %%%%%%%%%%%%%%%%%%%%%%%%%%%%%%%%%%%%%%%%%%%%%%%%%%%%%%%%%%%%%%%%%%%%%%%%%%%%%%%%
% %%%%%%%%%%%%%%%%%%%%%%%%%%%%%%%%%%%%%%%%%%%%%%%%%%%%%%%%%%%%%%%%%%%%%%%%%%%%%%%%

\begin{figure}
% \plotone{sec5_critical_stokes_ngvla.pdf}
\plotone{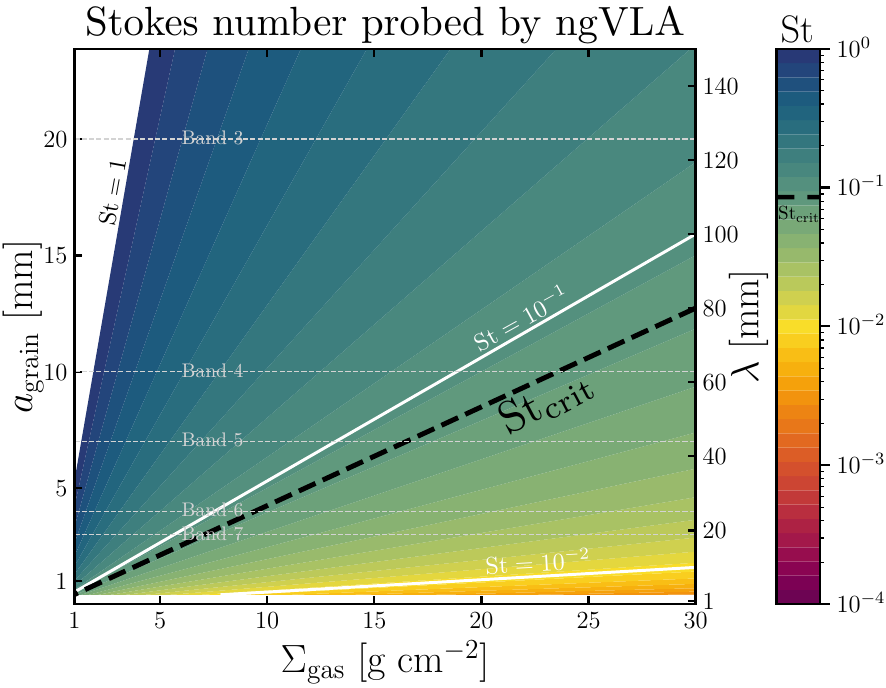}
\caption{ 
Dust Stokes number calculated for a range of gas surface densities and ngVLA dust grain sizes. For typical observing wavelengths of each ngVLA band, we mark the grain size probed assuming (LH y-axis) $\agrain = \lambdaobs/2\pi$ (RH y-axis). Long-wavelength observations with the ngVLA provide access to the $\St > \Stcrit$ regime.
\label{fig:sec5_intens-offset}}
\end{figure}

\textbf{Multi-wavelength observations.} Long-wavelength observations of dust spirals, for example with the ngVLA, %\textcolor{orange}{or with Band 1 of ALMA,} 
may allow us to probe larger, poorly-coupled ($\St > \Stcrit$) dust grains. In Figure \ref{fig:sec5_intens-offset}, we calculate the dust Stokes number of grain sizes relevant to the ngVLA ($\agrain \sim 1-20 \, \mm$) and overlay our estimate of the critical Stokes number, $\Stcrit$ (Eqn. \ref{eqn:Stcrit}), for a $1.0 \, \Mth$ planet at a disk radius $r= 1.7 \, \rp$, as found in \S\ref{subsec:dustgas-coupling}. The ALMA equivalent of this figure is Figure \ref{fig:sec3_stokes-azioffset-alma}(c). Comparing the spiral morphology in continuum observations of well-coupled vs. poorly-coupled dust opens new possibilities for future science.

Consider, for example, two observations of the same planet-driven dust spiral -- one obtained with ALMA Band 7 ($\agrain \approx 0.14 \, \mm$) and the second with ngVLA Band 5 ($\agrain \approx 7.0 \, \mm$). Assuming a local gas surface density of $\sim 3.0 \, \gcmsqrd$, these two observations would probe $\St = 10^{-2}$ (well-coupled) and $\St = 0.4$ (poorly-coupled) dust, respectively. These two species are shown in Figure \ref{fig:sec3_stokes-azioffset-alma}(a). At a distance of $0.7 \rp$ outside the planet, our hydrodynamic simulations predict an azimuthal offset between their spiral peaks in surface density of $20$ degrees. This is a significant offset that could feasibly be measured in observations. 

With a measurement of the azimuthal offset at a given disk radius, and with knowledge of its dependence on the Stokes number from fits to hydro simulations, one could estimate the underlying disk gas surface density, and subsequently measure the disk mass by repeating the exercise at different radii. This would be a direct evaluation of $\Sigg$ and $M_{\rm disk}$, independent from other methods, and free from the usual uncertainties that stem from making assumptions about the dust-to-gas ratio or dust opacity. We note however that to do this in practice requires taking into account temperature effects, which we discuss in more detail in \S\ref{sec:app:Tdust}. 

%%%%%%%%%%%%%%%%%%%%%%%%%%%%%%%%%%%%%%%%%%%%%%%%%%%%%%%%%%%%%%%%%%%%%%%%%%%%%%%%
%%%%%%%%%%%%%%%%%%%%%%%%%%%%%%%%%%%%%%%%%%%%%%%%%%%%%%%%%%%%%%%%%%%%%%%%%%%%%%%%

\textbf{Equation of state.} One of our main findings is that the observability of dust spirals depends heavily on how much the temperature spiral contributes to the overall spiral intensity, which in turn depends on the cooling timescale.
It would therefore be helpful if we had an idea of how quickly we expect disks to cool. 
As discussed in \S\ref{subsec:hydrosims}, typical values of $\tcool= \beta \, \Omega^{-1}(r)$ span a couple of orders of magnitude above and below unity at different radii within a single disk, and it is also likely that $\beta$ varies between disks (due to different dust properties, for example). More work is needed to constrain the cooling rate of individual target objects.

In addition to affecting the spiral detectability, the cooling timescale also muddles our ability to deduce the mass of the driving object in an observation, because the spiral intensity contrast is degenerate in $\Mp$ and $\beta$ (e.g. Fig. \ref{fig:sec3_Sig-T-Inu}, \ref{fig:sec3_spiral-arm-contrast} \& \ref{fig:app_multibeta}). A good knowledge on the cooling timescale in specific disks may also help distinguish the two spiral arm excitation mechanisms -- we find planet-driven spirals to be more prominent in continuum observations at \textit{longer} cooling timescales ($\beta \geq 10$), whereas GI-driven spirals have been found to express stronger velocity channel kinks at \textit{shorter} cooling timescales \cite[$\beta \leq 10$;][]{2021-longarini-privatecomm}.  

The planet-driven temperature spiral could also have consequences on the radial locations of icelines -- again, depending on $\beta$. The icelines of common molecules have been found not to correlate with observed locations of gaps \citep{2019-vandermarel-gaps}. Using radiative cooling, \citet{2020-ziampras-iceline} found that shock heating by the planet can raise the disk temperature high enough to displace the water iceline to larger radii. We find that a cooling rate of $\beta=10^{-1}$ or more is required for a $1.0 \, \Mth$ planet to drive a \textit{spiral} whose peak temperature is transiently $>10\%$ of the background (Fig. \ref{fig:app_multibeta} in \S\ref{sec:app:sec3-supplementary}), 
but that to get an azimuthally averaged \textit{ring} of 5\% temperature enhancement over the background outside the planet, a $\beta=10^{2}$ and a $3.0 \, \Mth$ planet on a fixed orbit are needed (not shown).

\textbf{Comparison to other published synthetic continuum images.} To our knowledge no previous theoretical works have specifically targeted the observability of planet-driven dust spirals in the ALMA continuum, but a few have provided synthetic ALMA continuum images of planet-hosting disks for alternative purposes -- and in some cases planet-driven dust spirals are visible in their results.

The image in Table 5 of \citet{2018-szulagyi-CPDalma} shows a model almost identical to one of our runs: a Band 7 continuum observation with a $1.0 \, \Mjup$ planet at $52 \, \au$ and measured rms noise of $15 \, \muJybeam$, though with the C50-28 configuration (i.e., assuming future antennas). As in our case (e.g. Fig. \ref{fig:sec4_multidim-table}), the outer primary spiral arm is directly visible in their image. They used the radiative 3D hydrodynamics code \texttt{JUPITER} (which includes heating by viscosity and adiabatic compression, and cooling by radiation and adiabatic expansion), and also emphasized that the temperature of one's target planet signature (in their case the CPD) influences its observability. They assumed, like us, $\Td = \Tg$, and unlike us, obtained their dust surface density distribution by scaling the gas (though for ALMA B7 this is a valid assumption; see Fig. \ref{fig:sec3_stokes-azioffset-alma}c).

\citet{2019-nazari} showed Band 7 observations achieving an rms noise of $18.5 \, \muJybeam$ for our medium angular resolution configuration pair, C43-5 + C43-8, of a disk with a low mass embedded planet  ($30 \, \Mearth$, equivalent to $90\%$ our $0.3 \, \Mth$ planet). They reported that no spirals were visible in their images. This agrees with our result in Fig. \ref{fig:sec4_lowmass} (with a slightly better rms noise of $13.6 \, \muJybeam$), where we showed that a residual map is needed to detect the spiral.

In residual maps of Band 6 observations of disks with a locally isothermal equation of state, spatially and temporally constant Stokes number, and two embedded $\geq 2.5 \, \Mth$ planets, \citet{2019-veronesi-diskmass} were able to detect the outer planet's inner primary and secondary spiral arm. In contrast to our parameter space, their outer planet was placed at $145 \, \au$.

\citet{2020-rowther-meru} investigated the influence of a migrating planet embedded in a gravitationally unstable disk in which %$\beta=\beta(r)$ 
$\beta$ varies radially with \texttt{PHANTOM} SPH simulations. They calculated their Band 6 ALMA continuum residuals using our same method, and found that spiral arms driven by an inwardly migrating $3.0 \, \Mjup$ planet initially at 160 au were visible in both face-on and $40^{\circ}$ inclined disks after $>1$ hr of integration time. They too found that a lower angular resolution observation, with its higher ${\rm S/N}$, allowed the planet-driven spirals to be seen more easily (see our Fig. \ref{fig:sec4_configuration}), though there they were comparing the C43-6 and C43-7 configurations unpaired. We provide B7 images obtained with individual configurations in Fig. \ref{fig:app_singleconfig}
% \ref{fig:app_singleconfig-1} \& \ref{fig:app_singleconfig-2-3} 
in \S\ref{sec:app:sec4-supplementary}.

\textbf{Caveats.} 
Our work can benefit from a few improvements to incorporate more realistic physics. We have restricted the planet to be on the simplest orbit, i.e., circular, co-planar, and non-migrating. Relaxing these assumptions may affect the morphology of the gap(s) \citep{2019-meru-insideout, 2019-nazari, 2019-perez-mininep, 2019-weber-migrating} and the spirals \citep{2003-quillen, 2015-duffell-chiang}, and impact the observability of spirals at a quantitative level. For example, \citet{2021-kanagawa-footprint} showed that the relative positions between the planet and the dust rings at gap edges depend on the migration rate of the planet. Meanwhile, the dust may be puffed up vertically in spirals \citep{krapp-2021}, which may result in detectable signatures in continuum observations \citep{2021-doi-kataoka}. It is impossible to capture such effects in our 2D simulations. In addition, the presence of multiple planets, as in the case of PDS 70 \citep{2019-haffert-PDS70}, may also complicate the recognition of individual spirals. 

We have ignored the effect of dust scattering, which can be important in optically thick disks \citep{2015-kataoka, 2019-zhu-scattering, 2019-liu}. While we mainly focus on optically thin to marginally optically thick cases, scattering may cause an order unity correction to the overall disk brightness when $\tau\sim1$ if the dust albedo is close to 1 \cite[Fig. 1 in][]{2020-sierra-lizano}, thus affecting the expected integration time to reach a desired S/N ratio. 
As a detailed sidenote, the increased dust surface density (thus $\tau$) locally at the spirals would introduce another correction factor, but we expect its impact on the spiral contrast to also be minimal as the spiral $\Sigd$ enhancements are only on the order of 10\%.

Finally, we adopt a single dust size most sensitively probed in observations with a fixed initial dust-to-gas mass ratio. In real systems, some dust mass is expected to be in grains of other sizes and thus does not contribute significantly to observations at a particular wavelength \citep{2018-birnstiel-dsharp5}. As such, the disk brightness in our models might be taken as upper (more optimistic) limits for our assumed initial dust-to-gas mass ratio (0.01).

%%%%%%%%%%%%%%%%%%%%%%%%%%%%%%%%%%%%%%%%%%%%%%%%%%%%%%%%%%%%%%%%%%%%%%%%%%%%%%%%
%%%%%%%%%%%%%%%%%%%%%%%%%%%%%%%%%%%%%%%%%%%%%%%%%%%%%%%%%%%%%%%%%%%%%%%%%%%%%%%%
\section{Summary \& Conclusions} \label{sec:conclusions}
%%%%%%%%%%%%%%%%%%%%%%%%%%%%%%%%%%%%%%%%%%%%%%%%%%%%%%%%%%%%%%%%%%%%%%%%%%%%%%%%
%%%%%%%%%%%%%%%%%%%%%%%%%%%%%%%%%%%%%%%%%%%%%%%%%%%%%%%%%%%%%%%%%%%%%%%%%%%%%%%%

Detecting a planet's spiral wake would constitute compelling evidence for its presence in the disk -- particularly if the inferred planet-spiral configuration is consistent with other signposts of the planet such as gaps/rings or local velocity kinks. 
In this work, we carry out 2D gas + dust hydrodynamic simulations and radiative transfer calculations. We produce synthetic Band 7 ALMA continuum observations of planet-driven dust spirals under a wide variety of disk and observing conditions. Our goal is to advise the search for planet-driven spirals in existing and future ALMA observations by identifying the most promising disk environments and observing specifications. We discuss the important physics underlying the observability of dust spirals in \S\ref{sec:results1} before presenting our simulated observations in \S\ref{sec:results2}. Our conclusions are as follows.

\begin{itemize}
    \item The ``critical'' Stokes number $\Stcrit$ dividing the well-coupled and poorly-coupled dust regimes can be estimated by equating the dust's intrinsic stopping time $\tstop$ with the gas spiral wake crossing time $\tcross$ (Eqn. \ref{eqn:tcross}).
    We find $\Stcrit \sim 0.05-0.1$ (Eqn. \ref{eqn:Stcrit}). Dust particles with $\St < \Stcrit$ form spirals identical to the driving gas spiral in morphology, while bigger particles azimuthally lag behind the gas peaks
    (Fig. \ref{fig:sec3_stokes-azioffset-alma}a,b), echoing \citet{2020-sturm}.
    At almost all gas surface densities and observing wavelengths, ALMA probes well-coupled ($\St < \Stcrit$) dust. Therefore, barring inclination or geometrical offsets, we expect dust spirals observed with ALMA to be excellent tracers of gas spirals at the midplane (Fig. \ref{fig:sec3_stokes-azioffset-alma}c). 
    \item 
    While the surface density contrast of well-coupled dust spirals depends non-monotonically on the cooling timescale $\beta$ and is the largest in locally isothermal disks, the strength of the temperature spiral formed in adiabatic disks increases monotonically with $\beta$. Adiabatic disks that cool slowly ($\beta \geq 10$) produce the hottest spirals with the largest contrast in surface brightness (Figs. \ref{fig:sec3_Sig-T-Inu} \& \ref{fig:app_multibeta}).
    \item The difference in brightness between dust spirals in slowly cooling vs. locally isothermal disks is most pronounced when the disk is optically thick (Fig. \ref{fig:sec3_spiral-arm-contrast}). 
    \item The signal of a spiral in a continuum image can be effectively highlighted in the residual map, enabling detections that otherwise may go unnoticed (Figs. \ref{fig:sec4_spread-Inu} \& \ref{fig:sec4_spread-SN}).
    \item %\textit{Ideal disk and planet conditions (\S\ref{subsec:alma-time}):}
    Planet-driven dust spirals are easiest to detect in adiabatic disks that cool slowly ($\beta \gtrsim 10$), that are marginally but not too optically thick ($\tauz \gtrsim 1.0$), and that host massive planets ($\Mp \gtrsim 1.0 \, \Mth$). In such disks, spirals can be detected with integration times on the order of hours (Fig. \ref{fig:sec4_multidim-table}).
    \item %\textit{Angular resolution (\S\ref{subsec:angular-resolution}):}
    Detecting a spiral is not contingent on resolving it. Higher angular resolution observations (beam size $\approx 0.5 \times$ the spiral width) have a higher spiral-signal-to-background ratio (they contain a greater proportion of spiral signal within each beam), but have a lower spiral-signal-to-\textit{noise} ratio. Lower angular resolution observations (beam size $\approx 2 \times$ the spiral width) ``wash out'' the spiral signal in the continuum image itself, but reveal the spiral with higher ${\rm S/N}$ in a residual map (Fig. \ref{fig:sec4_configuration}).
    \item %\textit{Low mass planets (\S\ref{subsec:lowmass-planets}):} 
    In a face-on, adiabatic ($\beta = 10$) marginally optically thick disk with a smooth dust surface density distribution exterior to the planet, we recover the outer spiral arm driven by a 1.6 Neptune mass planet ($0.3 \, \Mth$) around a solar-type star at $50 \, \au$ in the residuals of a Band 7 continuum observation obtained with an angular resolution of $62 \times 70\, \mas$ and a measured rms noise of $13.6 \,  \muJybeam$, achievable with 3.6 hours of on-source time and a full continuum bandwidth of 7.5 GHz (Fig. \ref{fig:sec4_lowmass}).
    \item %\textit{Gaps \& rings (\S\ref{subsec:gaps-rings}):} 
    The presence of gaps and rings can impair the observability of co-located spirals, by reducing the amount of disk area over which the spiral can be traced (Fig. \ref{fig:sec4_gaps-rings}). 
\end{itemize} 
Future continuum observations with the ngVLA may provide access to the poorly-coupled ($\St > \Stcrit$) dust spiral regime (Fig. \ref{fig:sec5_intens-offset}). Comparing the azimuthal location of dust spiral peaks in ALMA vs. ngVLA observations and measuring their offsets (Fig. \ref{fig:sec3_stokes-azioffset-alma}a) may enable direct constraints on the gas surface density and disk mass.

\begin{acknowledgments}
We are grateful to an anonymous referee for constructive suggestions that improved our paper. 
JS thanks: Pablo Ben\'itez-Llambay and Leonardo Krapp for maintaining and monitoring the public \texttt{FARGO3D} Bitbucket, and Shangjia Zhang and Dhruv Muley for providing simulations -- all of which aided in code comparisons; Ardjan Sturm for consultation where our works overlap; Jane Huang for helpful comments on an early version of the manuscript; Sarah Wood at the ALMA Help Desk and Nienke van der Marel for technical advice with the ALMA OT; Jeff Jennings and Brodie Norfolk for technical help on uvtables; and the Planet Formation Group at the University of Victoria for helpful discussions. 
Hydrodynamic simulations were performed on GPU computing nodes of Graham, B\'eluga and Cedar hosted by Compute Canada (www.computecanada.ca), as well as GASTRO hosted at McMaster University. 
RD and JS are supported by the Natural Sciences and Engineering Research Council of Canada (NSERC) and the Alfred P. Sloan Foundation. RAB was supported by a Royal Society University Research Fellowship, the STFC consolidated grant ST/S000623/1 and funding from the European Research Council (ERC) under the European Union's Horizon 2020 research and innovation programmes PEVAP (grant agreement number 853022) and DUSTBUSTERS (grant agreement number 823823). The National Radio Astronomy Observatory is a facility of the National Science Foundation operated under cooperative agreement by Associated Universities, Inc.
\end{acknowledgments}

\software{astropy \citep{astropy:2013, astropy:2018}, \texttt{CASA} \citep{2007-mcmullin-casa}, CMasher \citep{2020-cmasher}, matplotlib \citep{Hunter:2007}, numpy \citep{harris2020array}, pandas \citep{2011-mckinney-pandas}, scipy \citep{2020SciPy-NMeth} }

\dataavail{ Full set of synthetic ALMA observations and an image gallery are available at:\\ \href{https://doi.org/10.6084/m9.figshare.19148912}{https://doi.org/10.6084/m9.figshare.19148912}.}

%%%%%%%%%%%%%%%%%%%%%%%%%%%%%%%%%%%%%%%%%%%%%%%%%%%%%%%%%%%%%%%%%%%%%%%%%%%%%%%%
%%%%%%%%%%%%%%%%%%%%%%%%%%%%%%%%%%%%%%%%%%%%%%%%%%%%%%%%%%%%%%%%%%%%%%%%%%%%%%%%
\appendix

%%%%%%%%%%%%%%%%%%%%%%%%%%%%%%%%%%%%%%%%%%%%%%%%%%%%%%%%%%%%%%%%%%%%%%%%%%%%%%%%
%%%%%%%%%%%%%%%%%%%%%%%%%%%%%%%%%%%%%%%%%%%%%%%%%%%%%%%%%%%%%%%%%%%%%%%%%%%%%%%%
\section{Representative Disk Radial Profiles%\ref{sec:methods} 
%\section{Supplementary to \S2%\ref{sec:methods} 
} \label{sec:app:sec2-supplementary}

\begin{figure*}
\plotone{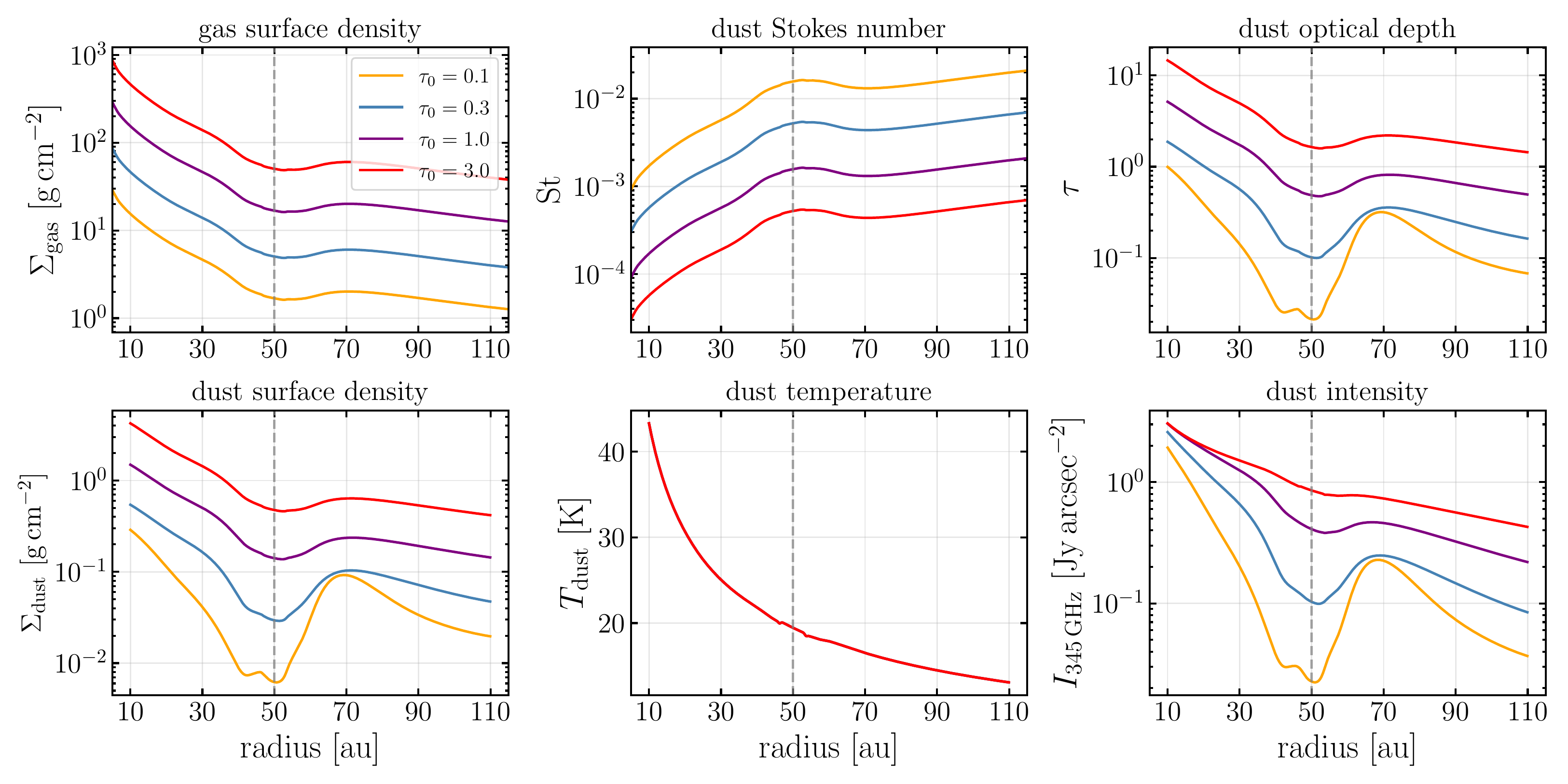}
\caption{Azimuthally averaged radial profiles of important disk quantities for four of our models, demonstrating the construction of the optical depth parameter, $\tauz$. Shown is the adiabatic ($\beta=10$) disk with a $1.0 \, \Mth$ embedded planet.
\label{fig:app_diskquantities}}
\end{figure*}

Figure \ref{fig:app_diskquantities} illustrates the differences between our $\tauz = 0.1$, $0.3$, $1.0$, and $3.0$ models by showing azimuthally averaged radial profiles of the relevant disk quantities (for one demonstrative permutation of EoS and $\Mp$). As described in \S\ref{subsec:radtranscalcs}, 
%we scale the gas surface density output from our hydrodynamic simulations 
four hydrodynamic simulations are run and normalized individually
such that the dust surface density (with an initial profile determined by a fixed dust-to-gas ratio of 0.01) gives an initial optical depth at $\rp$ of $0.1$, $0.3$, $1.0$, and $3.0$. By fixing the grain size to $\agrain = 0.14 \, \mm$, the Stokes number is different in each disk by the same factor as the gas surface density (\S\ref{subsec:gaps-rings}).

%%%%%%%%%%%%%%%%%%%%%%%%%%%%%%%%%%%%%%%%%%%%%%%%%%%%%%%%%%%%%%%%%%%%%%%%%%%%%%%%
\section{On assuming dust and gas thermal equilibrium at the midplane
} \label{sec:app:Tdust}

As described in  \S\ref{subsec:radtranscalcs}, we use the gas temperature output of our adiabatic \texttt{FARGO3D} hydrodynamic simulations as the dust temperature input for our radiative transfer calculations and subsequent ALMA observations of planet-driven dust spirals in adiabatic disks. Below, we justify this assumption and describe considerations for observers interested in longer wavelength observations (see discussion around Fig. \ref{fig:sec5_intens-offset} in \S\ref{sec:discussion}).

\textit{Under what conditions should the dust temperature $\Td$ be equal to the gas temperature $\Tg$?} The argument is simple: The temperature of a dust particle will be equal to that of its surroundings if it spends enough time in those surroundings to equilibrate. %This applies both inside the gas spiral wake and everywhere outside. 
We will show that the time it takes a dust particle to respond thermally happens to be very similar to the time it takes to respond aerodynamically, $\ttherm \approx \tstop$. Dust that is well coupled to the gas, as in our simulations, is therefore also well thermally coupled.

Consider a dust particle, with stopping time $\tstop$ and thermal coupling time $\ttherm$, flowing through a gas spiral wake with crossing time $\tcross$ (Eqn. \ref{eqn:tcross}). First, we express the particle's stopping time as
\begin{equation}
\tstop = \frac{\rhod \, \agrain}{\rhog \, \vth} \, ,
\end{equation}
where $\rhod$ and $\rhog$ are the dust and gas volume densities, $\agrain$ is the dust grain size and $\vth = \sqrt{8 \kB \Tg / \pi \mu \, m_{\rm H}}$ is the mean thermal speed of the gas molecules. As the particle enters the spiral wake, the temperature of its surroundings changes (e.g. Fig. \ref{fig:sec3_Sig-T-Inu}). The heating rate of a dust grain is given by \cite[e.g.,][]{1983-burke-hollenbach}
\begin{equation}
    m_{\rm dust} C_{\rm dust} \frac{\partial \Td}{\partial t} = \pi \agrain^2 n_{\rm gas} \vth \alpha \, (2 \kB \Tg - 2 \kB \Td) \, ,
\end{equation}
where $\alpha$ is the accommodation coefficient, $m_{\rm dust}$ and $C_{\rm dust}$ are the mass and specific heat capacity of the particle, $n_{\rm gas}$ is the gas number density and $\kB$ is Boltzmann's constant. This may be written as
\begin{equation}
    \frac{\partial \Td}{\partial t} = \Big[ \frac{4 \pi \agrain^2}{3 m_{\rm dust}} \rhog \vth \Big] \, \frac{3 \kB}{2 C_{\rm dust}} \alpha \, (\Tg - \Td) \, .
\end{equation}
Identifying the terms in square parentheses with $1/\tstop$ we arrive at
\begin{equation}
    \frac{\partial \Td}{\partial t} = \frac{1}{\tstop} \, \frac{3 \kB}{2 C_{\rm dust}} \, \alpha \, (\Tg - \Td) \, .
\end{equation}
We may thus define the thermal coupling time as
\begin{equation}
    \ttherm = \tstop \times \frac{2 C_{\rm dust}}{3 \kB} \, \alpha^{-1} \, .
    \label{eqn:ttherm-tstop}
\end{equation}

Since $\alpha \approx 1$ \citep{1983-burke-hollenbach}, the particle's thermal coupling time will be similar to its stopping time $\tstop \approx \ttherm$ if the specific heat capacity of the dust and gas are similar. Requiring enough time for a particle to come into thermal equilibrium with the gas inside the spiral wake, i.e. requiring $\ttherm < \tcross$, is therefore equivalent to requiring $\tstop < \tcross$, 
%We have shown that this is the case for dust with Stokes number 
which is our definition of being well coupled, $\St < \Stcrit$ (\S\ref{subsec:dustgas-coupling}). In other words, small dust grains probed by ALMA will have the same temperature as the gas. Their temperature spirals, and their intensity spirals, will be co-located.

The temperature spiral peaks of gas and poorly-coupled ($\St > \Stcrit$, and therefore $\ttherm > \tcross$) dust will \textit{not} be co-located. For observers interested in measuring the azimuthal offset between dust spirals in ALMA vs. ngVLA observations, this is a good thing; if the dust temperature peaks were aligned with that of the gas, the azimuthal offset between gas and large dust in the observed surface brightness would be reduced from that in surface density.
Whether the poorly-coupled dust temperature spiral peaks align with \textit{their own} surface density peaks requires future investigation, but it is promising that they are at least governed by similar timescales, $\ttherm \approx \tstop$.

%%%%%%%%%%%%%%%%%%%%%%%%%%%%%%%%%%%%%%%%%%%%%%%%%%%%%%%%%%%%%%%%%%%%%%%%%%%%%%%%
%%%%%%%%%%%%%%%%%%%%%%%%%%%%%%%%%%%%%%%%%%%%%%%%%%%%%%%%%%%%%%%%%%%%%%%%%%%%%%%%
\section{An alternative visualization of Figure 3%\ref{sec:results1} 
} \label{sec:app:sec3-supplementary}

\begin{figure*}
\plotone{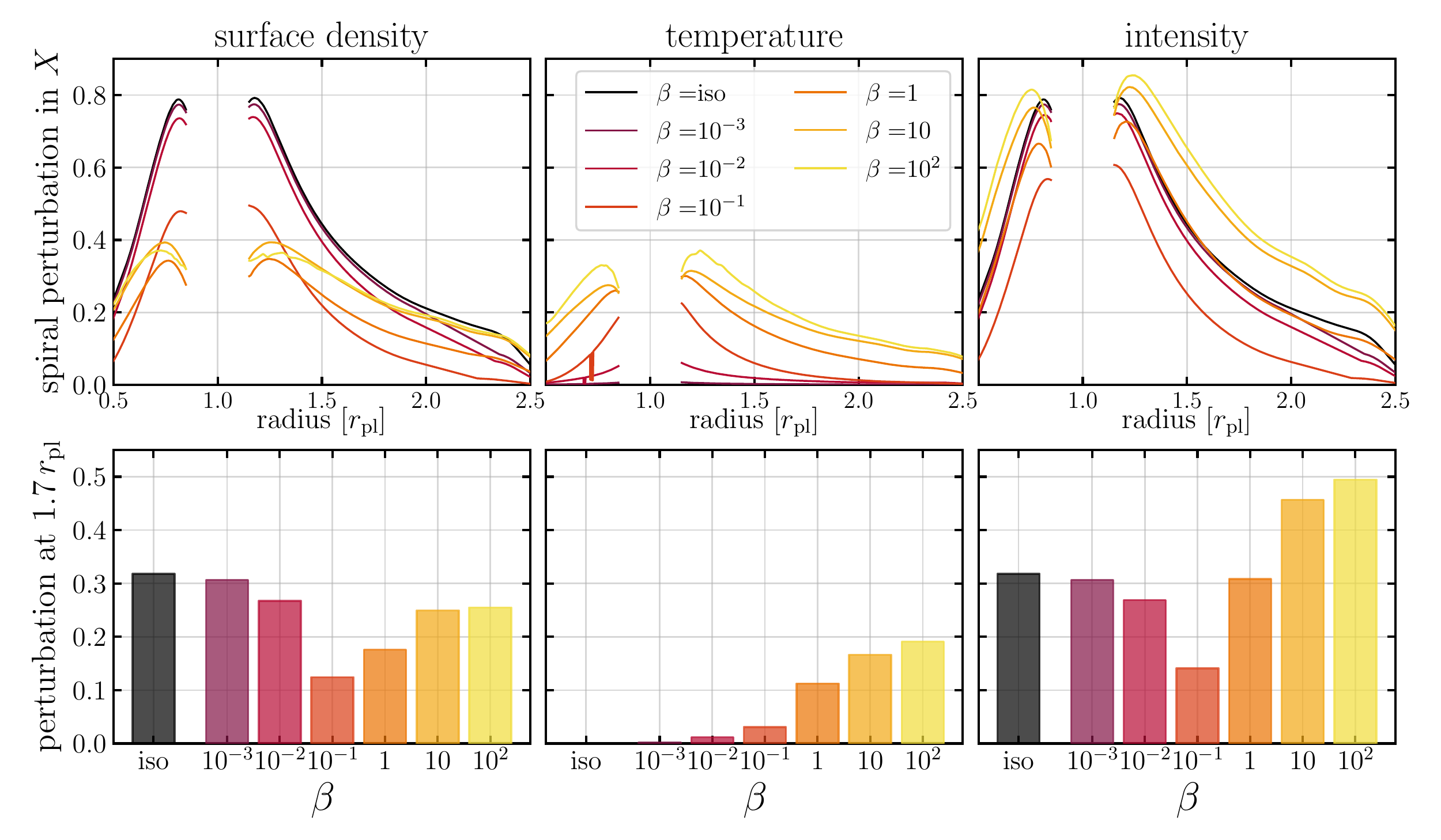}
\caption{ Like Figure \ref{fig:sec3_Sig-T-Inu}, but with the perturbations in each disk quantity traced along the inner and outer primary arms (\textbf{top panels}), and showing additional cooling timescales $\beta$. To highlight the non-monotonic dependence of the dust surface density and intensity perturbations on $\beta$, as well as to emphasize the spread in the amplitude generated by a given planet mass under different $\beta$, we show the values at $1.7 \, \rp$ as a bar chart (\textbf{bottom panels}). In this case the planet mass is $1.0 \, \Mth$. 
\label{fig:app_multibeta}}
\end{figure*}

Figure \ref{fig:app_multibeta} is an alternative visualization of Figure \ref{fig:sec3_Sig-T-Inu}, in which we trace the perturbation peaks in dust surface density, temperature and intensity (in the optically thin limit) along the inner and outer primary arms. We resolve the dependence of these quantities on the cooling timescale $\beta$ with additional values not shown in Fig. \ref{fig:sec3_Sig-T-Inu}. The amplitude of the perturbations varies substantially for the same $1.0 \, \Mth$ embedded planet.

%%%%%%%%%%%%%%%%%%%%%%%%%%%%%%%%%%%%%%%%%%%%%%%%%%%%%%%%%%%%%%%%%%%%%%%%%%%%%%%%
%%%%%%%%%%%%%%%%%%%%%%%%%%%%%%%%%%%%%%%%%%%%%%%%%%%%%%%%%%%%%%%%%%%%%%%%%%%%%%%%
\section{Synthetic observations of inclined disks} \label{sec:app:inclined-disks}

\begin{figure*}
\plotone{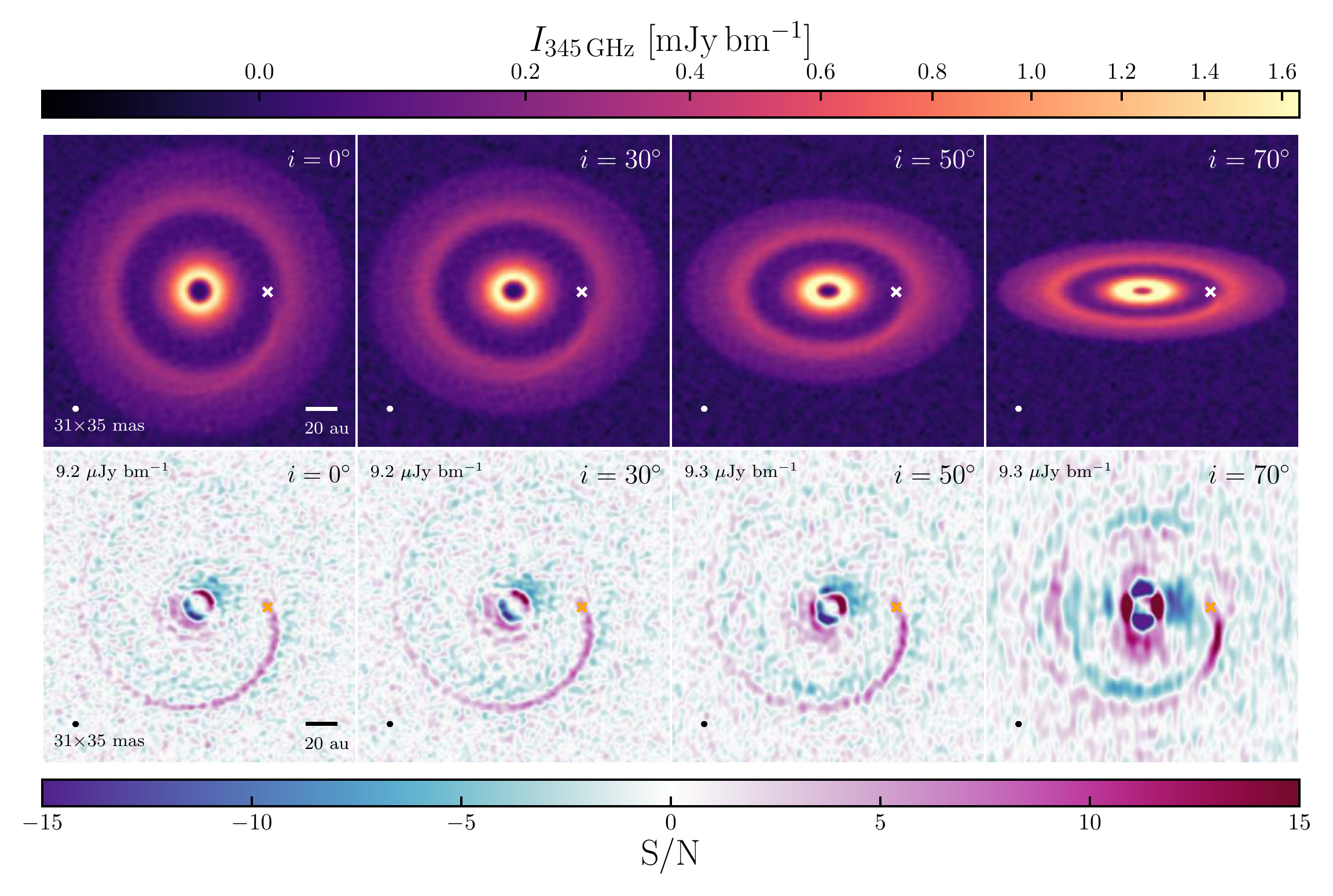}
\caption{Synthetic ALMA B7 continuum images (\textbf{top}) and deprojected residuals (\textbf{bottom}) of a demonstrative model disk, inclined by $30^{\circ}$, $50^{\circ}$ and $70^{\circ}$ to compare spiral visibility under varying disk inclination. The effect of inclination is not significant unless the disk is very inclined (i.e. $70^{\circ}$, rightmost column).
\label{fig:app_inclined}}
\end{figure*}

As a first start in informing observations of inclined disk systems, we generate an additional set of continuum images of tilted disks. The underlying dust surface density and temperature maps have been ``squished'' along the north-south axis by a factor of $\cos{(i)}$, and the dust surface density scaled by a factor of $1/\cos{(i)}$, before the emergent dust intensity was calculated. The optical depths of the inclined disks are thus different to their face-on counterparts, but we still use the $\tauz$ parameter to refer to them.

We also experimented with changing the position angle of the planet by 90 degrees (such that it was located on the north-south axis) and found that the spiral visibility in the deprojected residual maps was not greatly affected.

The images in Figure \ref{fig:app_inclined} depict a $\Mp = 1.0 \, \Mth$ planet embedded in an optically thin ($\tauz = 0.1$) disk with an adiabatic equation of state ($\beta=10$), observed with the C43-5 + C43-8 configuration pair for a combined on-source time of 8.02 hours. The synthesized beam is shown in the bottom left corner of each image, and the measured rms noise is shown in the top left corner of the deprojected residual panels. View all our model permutations at \href{https://doi.org/10.6084/m9.figshare.19148912}{https://doi.org/10.6084/m9.figshare.19148912}.

%%%%%%%%%%%%%%%%%%%%%%%%%%%%%%%%%%%%%%%%%%%%%%%%%%%%%%%%%%%%%%%%%%%%%%%%%%%%%%%%
%%%%%%%%%%%%%%%%%%%%%%%%%%%%%%%%%%%%%%%%%%%%%%%%%%%%%%%%%%%%%%%%%%%%%%%%%%%%%%%%
\section{Considerations behind synthetic observations
%\section{Supplementary to \S4%\ref{sec:results2} 
} \label{sec:app:sec4-supplementary}

%%%%%%%%%%%%%%%%%%%%%%%%%%%%%%%%%%%%%%%%%%%%%%%%%%%%%%%%%%%%%%%%%%%%%%%%%%%%%%%%
\begin{deluxetable}{ccccc} 
\tablenum{1}
\tablecaption{ALMA observing time for requested sensitivities. \label{tab:sensitivity-obstime} }
\tablehead{
\colhead{Requested} &
\colhead{Compact} & 
\colhead{Extended}  & 
\twocolhead{Combined}  \\
\colhead{Sensitivity} &
\colhead{On-source} &
\colhead{On-source} &
\colhead{OS} &
\colhead{OS+OH} \\
\colhead{($\muJybeam$)} &
\colhead{(hr)} &
\colhead{(hr)} &
\colhead{(hr)} &
\colhead{(hr)} 
}
\startdata
             & C43-4     & C43-7     &           &                   \\\hline
10           & 1.51      & 6.57      & 8.08      & 18.51             \\
15           & 0.67      & 2.92      & 3.59      & 8.35              \\
20           & 0.38      & 1.64      & 2.02      & 4.63              \\
25           & 0.24      & 1.05      & 1.29      & 3.15              \\
30           & 0.17      & 0.73      & 0.90      & 2.09              \\
35           & 0.12      & 0.53      & 0.66      & 1.62              \\\hline
             & C43-5     & C43-8     &           &                   \\\hline
10           & 1.45      & 6.57      & 8.02      & 21.94             \\
15           & 0.64      & 2.92      & 3.56      & 10.08             \\
20           & 0.36      & 1.64      & 2.00      & 5.49              \\
25           & 0.23      & 1.05      & 1.28      & 4.03              \\
30           & 0.16      & 0.73      & 0.89      & 2.52              \\
35           & 0.12      & 0.53      & 0.65      & 2.08              \\\hline
             & C43-6     & C43-9     &           &                   \\\hline
10           & 1.38      & 6.57      & 7.95      & 18.21             \\
15           & 0.61      & 2.92      & 3.53      & 8.22              \\
20           & 0.34      & 1.64      & 1.98      & 4.55              \\
25           & 0.22      & 1.05      & 1.27      & 3.10              \\
30           & 0.15      & 0.73      & 0.88      & 2.06              \\
35           & 0.11      & 0.53      & 0.65      & 1.60              \\
\enddata
\tablecomments{Since the C43-6 + C43-9 configuration pair was not available in Cycle 8, the Cycle 7 ALMA OT was used to determine the observing times for that pair.}
\end{deluxetable}
%%%%%%%%%%%%%%%%%%%%%%%%%%%%%%%%%%%%%%%%%%%%%%%%%%%%%%%%%%%%%%%%%%%%%%%%%%%%%%%%

\begin{figure} 
\plotone{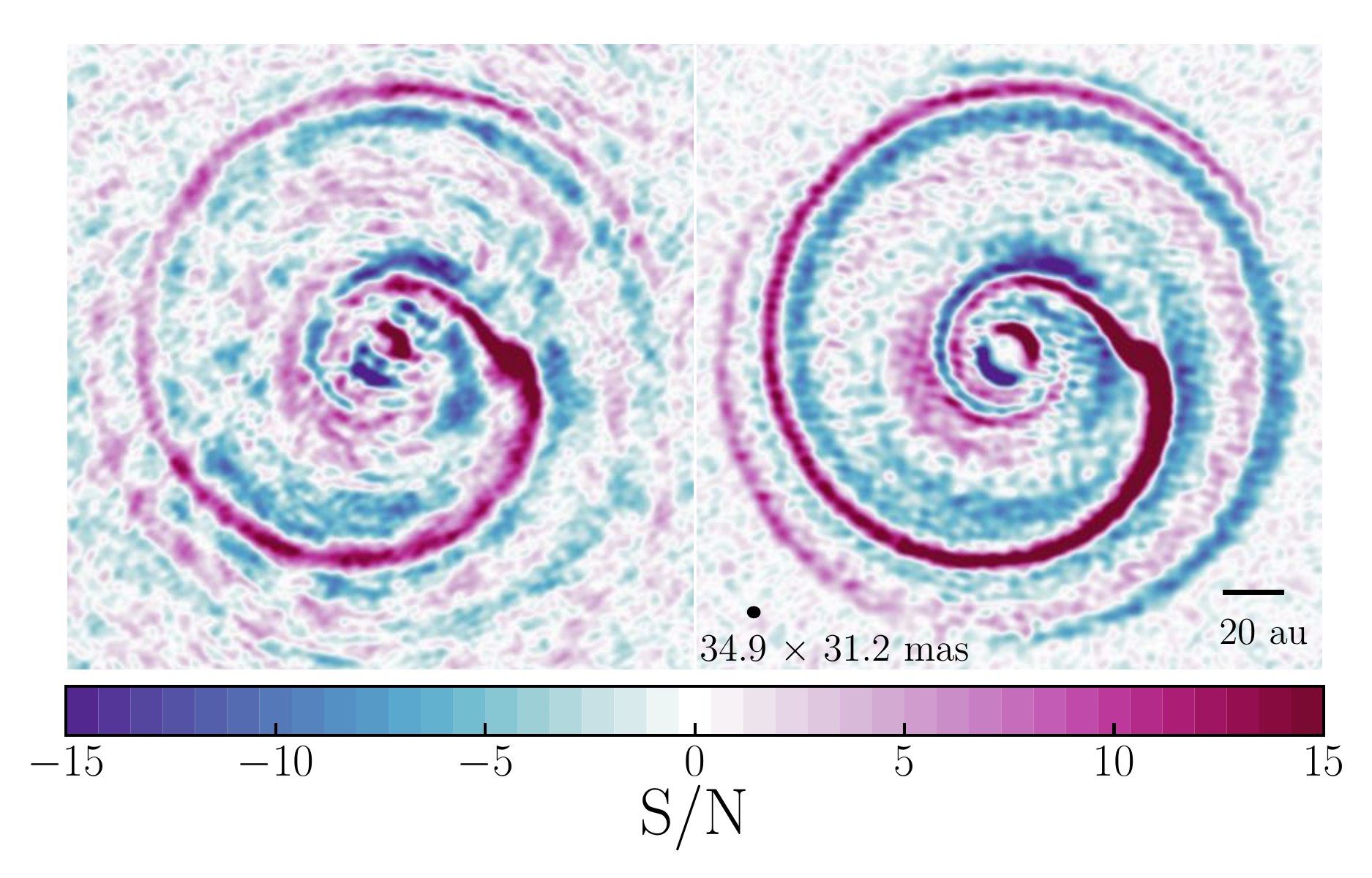}
\caption{(\textbf{Left}) Imaged \texttt{frank} visibility residuals generated with parameters that best revealed the planet-driven spiral of the parameters that we explored. (\textbf{Right}) Residuals generated with the method used in this work. The measurement set shared between these two images corresponds to the model that was presented in panel (h) of Figs. \ref{fig:sec4_spread-Inu} \& \ref{fig:sec4_spread-SN} and represents one of the strongest spiral recoveries out of our full set of 432 model images.
\label{fig:app_frank}}
\end{figure}

\begin{figure*}
\plotone{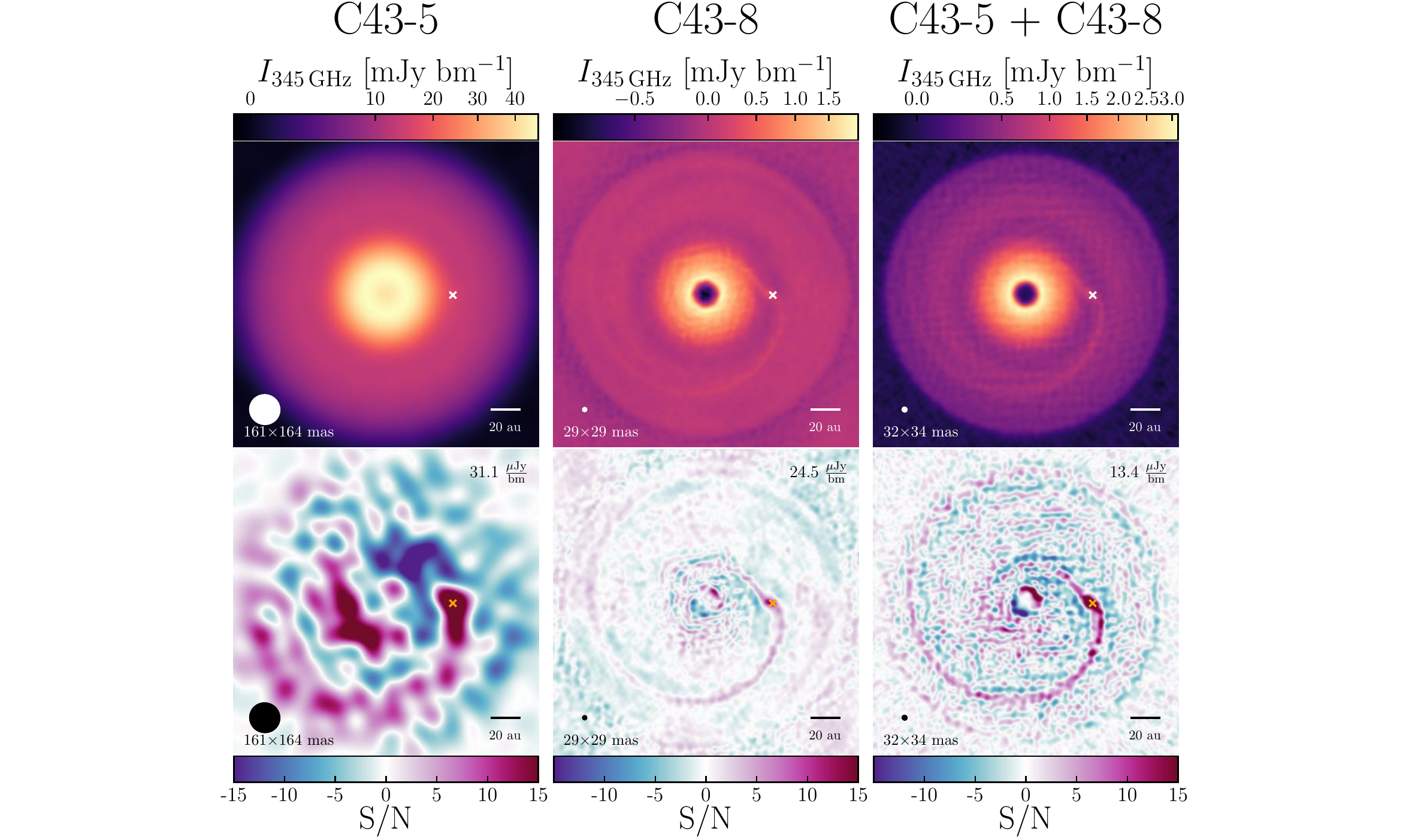}
\caption{Synthetic ALMA B7 continuum observations with the compact and extended configurations that correspond to the pair shown in the middle column of Fig. \ref{fig:sec4_configuration}. Following the Sensitivity Calculator in the ALMA OT, the combined on-source time of 3.56 hrs for this requested sensitivity of $15 \, \muJybeam$ was distributed as 0.64 hrs with C43-5 and 2.92 hrs with C43-8 (see also Table \ref{tab:sensitivity-obstime}). The measured rms noise in each observation is written in the top right corner of the residual maps.
\label{fig:app_singleconfig}}
\end{figure*}

In Table \ref{tab:sensitivity-obstime} we provide the individual and combined on-source (OS) times used in this work, which were specified by the ALMA Cycle 8 OT as what is needed to achieve each selected requested continuum sensitivity with each configuration pair at 345 GHz for a bandwidth of 7.5 GHz. For context, we also show the total observing time including overheads (OT), which does not scale linearly with the estimated time on source. Our highest sensitivity observation requires 8 SB executions.

As described in \S\ref{sec:results2}, we experimented with using \texttt{frank} \citep{2020-jennings-frank} to create residual maps and highlight the planet-driven spiral signal in our synthetic continuum observations. We fitted the observed visibilities with \texttt{frank} (exploring 16 permutations of the hyperparameters $w_{\rm smooth}$ and $\alpha$ and using the corrected weights), converted the residual uvtable into a measurement set, and imaged that measurement set with \texttt{tclean} in a way identical to as was done for the synthetic continuum observations, with the exception of setting the number of iterations to zero. Figure \ref{fig:app_frank} provides a comparison between \texttt{frank} and our method of calculating residuals directly in the image plane (e.g. Fig. \ref{fig:sec4_spread-SN}). We found that with the corrected weights, the visibility residuals were most similar to our image plane residuals, and were very insensitive to $w_{\rm smooth}$ and $\alpha$. Figure shows \ref{fig:app_frank} the results with parameters $w_{\rm smooth}=1.01$, $\alpha=1.05$ and $R_{\max} = 1.2 \arcsec$.

In Figure \ref{fig:app_singleconfig} 
we show continuum observations made from the compact and extended configuration measurement sets that were concatenated to create the continuum images presented in the middle column of Figure \ref{fig:sec4_configuration}. The individual measurement sets were imaged by the same procedure as was their combination. The residual maps of the extended configuration observation demonstrate the long baseline artifacts mentioned in \S\ref{subsec:almaobs}. These artifacts can be generally characterized as repeating patterns of large regions on the sky with over- or under-intensity, in the rough shape of stripes, slices of pie, or wide spokes, depending on the configuration. 

\bibliography{manuscript}{}
\bibliographystyle{aasjournal}

\end{document}